\documentclass[11pt,a4paper]{article}
\usepackage{graphicx}
\usepackage{jheppub}
\usepackage{subfig}
\usepackage[latin1]{inputenc}
\usepackage{amsmath}
\usepackage{amsfonts}
\usepackage{amssymb}
\usepackage{setspace}
\usepackage{mathtools}
\usepackage{pstricks}
\usepackage{color}

\providecommand{\f}[2]{\frac{{#1}}{{#2}}}

\newcommand{\dda}{\ensuremath{\ddot{a}}}

\newcommand{\ba}{\begin{eqnarray}}
\newcommand{\ea}{\end{eqnarray}}

\title{Quantum Corrections to Inflaton and Curvaton Dynamics}

\author[a]{Tommi Markkanen,}
\author[b]{Anders Tranberg}

\affiliation[a]{Helsinki Institute of Physics and Department of Physics, P. O. Box 64, FI-00014, University of Helsinki, Finland.}
\affiliation[b]{Niels Bohr International Academy and Discovery Center, Niels Bohr Institute,\\  Blegdamsvej 17, 2100 Copenhagen, Denmark}

\emailAdd{tommi.markkanen@helsinki.fi}
\emailAdd{anders.tranberg@nbi.dk}

\abstract{We compute the fully renormalized one-loop effective action for two interacting and self-interacting scalar fields in FRW space-time. We then derive and solve the quantum corrected equations of motion both for fields that dominate the energy density (such as an inflaton) and fields that do not (such as a subdominant curvaton). In particular, we introduce quantum corrected Friedmann equations that determine the evolution of the scale factor. We find that in general, gravitational corrections are negligible for the field dynamics. For the curvaton-type fields this leaves only the effect of the flat-space Coleman-Weinberg-type effective potential, and we find that these can be significant. For the inflaton case, both the corrections to the potential and the Friedmann equations can lead to behaviour very different from the classical evolution. Even to the point that inflation, although present  at tree level, can be absent at one-loop order.}


\begin{document}

\maketitle

\section{Introduction}
\label{sec:introduction}

High-energy processes in the early Universe must ultimately be described in terms of interacting quantum fields evolving in a curved space-time background. Often the dynamics is described by classical field equations of motion and the Friedmann equations for the scale factor, which are solved analytically or numerically. A plethora of models exist, where scalar field potentials are designed to generate a certain behaviour and certain physical phenomena. Although rarely stated explicitly, the understanding is that the {\it classical} field is really the homogeneous one-point function, or mean field, of a quantum field operator, rolling in a quantum effective potential. This connection is often left unclear, and it is not obvious that for a given effective potential, there exists such an underlying renormalizable theory. In fact, consistent renormalization is rarely addressed in this context, opening up a number of pitfalls.

In recent years, interacting quantum fields have become the subject of renewed interest in the general context of inflation, with the advent of searches for non-gaussianity in the Cosmic Microwave Background, curvaton models \cite{curvaton,Lyth}, preheating \cite{preheating} and dissipative inflation models \cite{warminflation}. A number of approaches and approximations to the quantum dynamics have been employed, often motivated by practical tractability and features of the specific problem allowing for certain simplifications. In many cases, quantum corrections of one type or another are added to the classical dynamics, while others are neglected. It is not always clear that such choices are systematic, i.e. in terms of an improvable sequence of approximations.

Quantum field theory in curved space-time has a long history (see \cite{ParkerToms,birreldavies} and references therein) and since it is a first principle method, the difficulties in cosmological perturbation theory related to renormalization \cite{Lyth:2012yp} and choosing the correct sized box for the curvature perturbations \cite{Lyth:2007jh}, are not present. Current work includes \cite{Barvinsky:1998rn, Bezrukov:2007ep, DeSimone:2008ei, Barvinsky:2008ia, Bezrukov:2010jz}. There has also recently been a focus on applications of the CTP ("closed-time-path") or Schwinger-Keldysh ("in-in") formalism to such curved spaces. In particular, the 2PI ("two-particle irreducible") effective action formalism was pioneered in this context by Calzetta and Hu \cite{calzettahu} (see also \cite{Ramsey:1997sa}). This is a very powerful framework, where evolution equations for the mean field and the propagator are solved self-consistently, resumming a large class of perturbative diagrams. The effects of quantized gravity have also been studied and provide interesting modifications to the inflation paradigm \cite{Tsamis:1996qq,Tsamis:1996qm,Romania:2012av} and references therein.

A drawback is that beyond leading order the 2PI equations are numerically very hard to solve. Apart from \cite{tranberg}, work has therefore concentrated on the leading order approximations, often in strict deSitter space \cite{prokopec,sloth,serreau} but also for self-consistent FRW Universes following from the Friedmann equation \cite{Ramsey:1997sa,Baacke:1999gc,Baacke:2010bm}  (see also \cite{boyanovsky, Dimopoulos:2003ss}). These approximations are Gaussian in the sense that the only non-zero connected correlators are the mean field and the propagator\footnote{It is however possible to obtain non-Gaussian curvature perturbations from Gaussian field perturbations.}. Gaussian dynamics possesses a non-physical fixed point, and does not allow for dissipation and thermalization (see for instance \cite{fixpoint}).

The issue of renormalization in this context has also received abundant attention. The benchmark method is adiabatic regularization \cite{adiabatic}, see also \cite{ParkerToms,birreldavies}, where counterterms are computed typically at the level of the evolution equations for the interacting fields; but also to renormalize the quantum expectation value of the energy-momentum tensor in the semi-classical Friedmann equation \cite{Ramsey:1997sa}. (For a different renormalization method functioning at the level of the evolution equations see \cite{Baacke:1999gc,Baacke:2010bm}). Adiabatic regularization is a very elegant and intuitive approach, whereby computing the adiabatic vacuum to fourth order in derivatives of the scale factor, all divergences can be identified and subtracted. One must then have some prescription to fix the finite parts of the counterterms. 

However, it is possible to side-step some of these issues, if one is primarily interested in the quantum corrections to the mean field. This applies to the case where the field in question evolves in a given FRW background (we shall refer to this as a spectator or {\it curvaton} field), and to the case where the field itself dominates the energy density (we call this a participating or {\it inflaton} field), and where we therefore need to solve a quantum corrected Friedmann equation in addition to the mean field equation of motion. For this purpose, it is sufficient to consider the 1PI effective action. 

In this paper, we compute the fully renormalized one-loop 1PI effective action for a two-scalar theory. 
This allows us to study inflaton and curvaton evolution coupled to another field, and the effect of quantum fluctuations on the dynamics. As an alternative to adiabatic regularization, which was used for similar problems in \cite{MolinaParis:2000zz,Anderson:2008dg}, we opt to renormalize at the level of the action using an expansion of the heat kernel \cite{Hu:1984js,Kirsten:1993jn,Elizalde:1994ds}. This has the advantage that we can impose a set of renormalization conditions, which allow us to fix the counterterms completely. From the renormalized (and therefore finite) effective action, we derive evolution equations for both the mean fields and the metric by variation. These we solve numerically.

The paper is organized as follows: In section  \ref{sec:effeaction} we write down the effective action. At this point we will simply state the result, which in the most general version includes interactions between the fields as well as self-interactions, and allows for non-minimal coupling to gravity.  To illustrate the main effects of quantum corrections, we then immediately specialize to two simpler cases for which we solve the evolution equations in section \ref{sec:dynamics}; first a curvaton in a quadratic potential coupled to another field in its vacuum; then a quadratic inflaton also coupled to such a field. We do not restrict ourselves to deSitter space.

In section \ref{sec:computation} we present in complete detail how we arrived at the effective action stated in section \ref{sec:effeaction} including the crucial issue of renormalization, as well as a discussion of the approximations made and the limitation that apply to the result. Further technical points can be found in the appendices. We conclude in section \ref{sec:conclusions}.

\section{Effective action of two-scalar field model}
\label{sec:effeaction}
We consider two real scalar fields $\sigma$ and $\phi$ in what will eventually be a flat FRW background, but which for the moment is described by a general metric $g^{\mu\nu}$. Depending on the context, the fields could be the inflaton, the curvaton or some other scalar fields, but at this point they are general and on equal footing. Our convention for the metric is $(-,+,+,+)$ and the Riemann and Einstein tensors\footnote{${R^\mu}_{\alpha \beta \gamma} =  \Gamma^\mu_{\alpha \gamma,\beta}-\Gamma^\mu_{\alpha \beta,\gamma}+\Gamma^\mu_{\sigma \beta}\Gamma^\sigma_{\gamma \alpha}-\Gamma^\mu_{\sigma \gamma}\Gamma^\sigma_{\beta \alpha}$, $G_{\mu\nu}=-\f{1}{2} Rg_{\mu\nu}+R_{\mu\nu}$.} are defined according to the $"+"$ conventions of \cite{Misner:1974qy}. The bare (classical) action reads
\ba
\label{eq:treelevel}
S[\phi,\sigma,g^{\mu\nu}]\equiv S_m[\phi,\sigma,g^{\mu\nu}]+S_g[g^{\mu\nu}],
\ea
where
\ba
\label{eq:matteraction}
S_m[\phi,\sigma,g^{\mu\nu}]&\equiv& \int d^4x\sqrt{-g}~\bigg[-\frac{1}{2}g^{\mu\nu}\partial_\mu\phi\partial_\nu\phi+\eta_\phi\Box\phi^2-\frac{m_\phi^2}{2}\phi^2-\f{1}{2}(1-\xi_\phi)\f{R}{6}\phi^2\nonumber \\
&&
-\frac{1}{2}g^{\mu\nu}\partial_\mu\sigma\partial_\nu\sigma+\eta_\sigma\Box\sigma^2-\frac{m_\sigma^2}{2}\sigma^2-\f{1}{2}(1-\xi_\sigma)\f{R}{6}\sigma^2\nonumber \\
&&-\f{g\phi^2\sigma^2}{4}-\f{\lambda_\sigma \sigma^4}{4!}-\f{\lambda_\phi\phi^4}{4!}\bigg],
\ea
and
\ba
S_g[g^{\mu\nu}]\equiv \int d^4x\sqrt{-g}~\bigg[\Lambda+\alpha R+\beta R^2+\epsilon_1 C^2+\epsilon_2 G+\kappa \Box R\bigg],
\ea
We have introduced the square of the Weyl tensor $C^2$ and $G$ is the Gauss-Bonnet density, defined as
\ba
\label{eq:CG}
C^2 =\frac{1}{3}R^2-2R^{\mu\nu}R_{\mu\nu}+R^{\mu\nu\rho\sigma}R_{\mu\nu\rho\sigma},\qquad G =R^2-4R^{\mu\nu}R_{\mu\nu}+R^{\mu\nu\rho\sigma}R_{\mu\nu\rho\sigma},
\ea
in terms of the Ricci scalar $R$, the Ricci tensor $R^{\mu\nu}$ and the Riemann tensor $R^{\mu\nu\rho\sigma}$. We also have the operator $\Box=\nabla_\mu\nabla^\mu=|g|^{-1/2}\partial_\mu(|g|^{1/2}\partial^\mu)$. We assume that $m_\phi^2$, $m_\sigma^2$, $\lambda_\phi$, $\lambda_\sigma$ and $g$ are all positive. Negative masses squared could be considered, leading to spontaneous symmetry breaking, which is especially interesting in curved space \cite{Shore:1979as}, but we will refrain from doing so here. The couplings $\eta$ and $\kappa$ multiply total divergences and in unbounded spaces the boundary terms vanish. We also used the parametrization for the $\xi$ for which $\xi=0$ or 1 means the conformal or minimal coupling respectively.
$\beta=\epsilon_1=\epsilon_2=\kappa=0$ and $\alpha=1/16\pi G_N$ ($G_N$ is the Newton constant) is the Einstein-Hilbert action with two scalar fields. In terms of the ``usual'' cosmological constant $\Lambda'$, we have $\Lambda'=-8\pi G_N \Lambda$. To each of the 15 parameters corresponds a counterterm, which will come into play below, when we discuss renormalization. 

The 1PI effective action $\Gamma[\phi,\sigma,g^{\mu\nu}]$ is a functional of the mean fields\footnote{We will use the same symbols $\sigma$, $\phi$ to denote the mean fields. It will be clear from the context whether we speak about the mean fields or the field operators $\sigma(x,t)$, $\phi(x,t)$.}
\ba
\sigma(t)=\langle\sigma(x,t)\rangle,\qquad \phi(t)=\langle\phi(x,t)\rangle,
\ea
which for a homogeneous state have only time dependence. We will treat $g^{\mu\nu}$ as a classical external field. There are therefore no quantum gravity corrections included in our treatment, only corrections from loops of the scalar fields, which as we will see include gravitational operators. Once the effective action is known, it is straightforward to derive the equations of motion by variation
\ba\label{eq:EOM}
\frac{\delta \Gamma[\phi,\sigma,g^{\mu\nu}]}{\delta g^{\mu\nu}}=0,\quad\frac{\delta \Gamma[\phi,\sigma,g^{\mu\nu}]}{\delta\phi}=0,\quad\frac{\delta \Gamma[\phi,\sigma,g^{\mu\nu}]}{\delta\sigma}=0.
\ea
The first of these amounts to a quantum corrected Einstein equation, and the last two are quantum corrected field equations of motion.

We compute $\Gamma$ at one loop order, and fix the counterterms in such a way that at the point $g^{\mu\nu}=\eta^{\mu\nu}$, $\sigma=\phi=0$ the renormalized couplings are equal to the tree-level ones. We postpone the details of the computation and renormalization of the effective action to section \ref{sec:computation}. Truncating a derivative expansion in the way described there, we have the general result
\ba
\Gamma[\phi,\sigma,g^{\mu\nu}] = \int d^4x \sqrt{-g}\,\mathcal{L}_{eff}=\int d^4x\sqrt{-g}\,\left[\mathcal{L}_{eff}^{(0)}+\mathcal{L}_{eff}^{(1)}\right],
\ea
with a part that looks like the tree-level action (but now in terms of mean fields)
\ba
\label{main1}
\mathcal{L}_{eff}^{(0)}&=&
-\frac{1}{2}g^{\mu\nu}\partial_\mu\phi\partial_\nu\phi-\frac{m_\phi^2}{2}\phi^2-\f{1}{2}(1-\xi_\phi)\f{R}{6}\phi^2-\f{\lambda_\phi\phi^4}{4!}\nonumber \\
&&
-\frac{1}{2}g^{\mu\nu}\partial_\mu\sigma\partial_\nu\sigma-\frac{m_\sigma^2}{2}\sigma^2-\f{1}{2}(1-\xi_\sigma)\f{R}{6}\sigma^2-\f{\lambda_\sigma \sigma^4}{4!}-\f{g\phi^2\sigma^2}{4}\nonumber \\
&&+\Lambda+\alpha R+\beta  R^2+G \epsilon_2,
\ea
and a quantum correction
\ba
\label{main2}
\mathcal{L}_{eff}^{(1)}&=&\f{1}{64\pi^2}\bigg\{\f{1}{24}\Big[6 g \left(g+3 \left(\lambda _{\sigma }+\lambda _{\phi }\right)\right)\sigma ^2 \phi ^2 + \left(\xi _{\sigma }^2+\xi _{\phi }^2\right)R^2-4  \left(\xi _{\sigma } m_{\sigma }^2+m_{\phi }^2 \xi _{\phi }\right)R\nonumber\\
&&
+9 \left(g^2+\lambda _{\phi }^2\right) \phi ^4+12 \left(g m_{\sigma }^2+m_{\phi }^2 \lambda _{\phi }\right) \phi ^2-6  \left(g \xi _{\sigma }+\lambda _{\phi } \xi _{\phi }\right)R \phi^2\nonumber \\
&&
+9  \left(g^2+\lambda _{\sigma }^2\right)\sigma^4+12 \left(gm_\phi^2+ m_{\sigma }^2\lambda _{\sigma }\right)\sigma^2-6  \left(g \xi _{\phi}+\lambda _{\sigma } \xi _{\sigma }\right)R \sigma ^2
\Big]\nonumber \\
&&
+\bigg[\frac{G}{360}  -\frac{g^2 \sigma ^2 \phi ^2 m_{\phi }^2}{4 (m_{\phi }^2-m_{\sigma }^2)}
-\f{1}{2}\left(m_{\phi }^2+\frac{\phi ^2 \lambda _{\phi }}{2}+\frac{g \sigma ^2}{2}-\frac{R \xi _{\phi }}{6}\right)^2\bigg]\log \bigg(\frac{M_-^2 M_+^2}{m_{\phi }^4}\bigg)\nonumber \\
&&
+ \bigg[\frac{G}{360} +\frac{g^2 \sigma ^2 \phi ^2 m_{\sigma }^2}{4 (m_{\phi }^2-m_{\sigma }^2)}
-\f{1}{2}\left(m_{\sigma }^2+\frac{\sigma ^2 \lambda _{\sigma }}{2}+\frac{g \phi ^2}{2}-\frac{R \xi _{\sigma }}{6}\right)^2\bigg]\log \bigg(\frac{M_-^2 M_+^2}{m_{\sigma }^4}\bigg)
\nonumber \\
&&
+\frac{1}{12} \log \bigg(\frac{M_-^2}{M_+^2}\bigg) \Big[3  \sigma ^2(g+\lambda_\sigma)+3 \phi ^2(g+\lambda_\phi)
+6 m_{\sigma }^2+6 m_{\phi }^2- R(\xi_\sigma+\xi_\phi)
\Big]
(M_+-M_-)\bigg\}.\nonumber\\
\ea
We have defined
\begin{equation}
M^2_\pm=\f{m_\phi^2+m_\sigma^2}{2}-\f{\xi_\phi+\xi_\sigma}{2}\f{R}{6}+\f{\Phi^2_\pm}{2},
\end{equation}
where the mass-eigenstate fields are given by
\ba
\label{eq:eigenf}
\Phi^2_\pm&=&\frac{1}{2} g \left({\sigma}^2+{\phi^2}\right)+\f{\lambda_\phi{\phi}^2+\lambda_\sigma{\sigma}^2}{2} \pm\bigg[\bigg(m_{\phi }^2-m_{\sigma }^2+\frac{1}{6} R \left(\xi _{\sigma }-\xi _{\phi }\right)\bigg)^2\nonumber \\
&&+\left(m_{\phi }^2-m_{\sigma }^2+\frac{1}{6} R \left(\xi _{\sigma}-\xi _{\phi}\right)\right)\left({\sigma} ^2 \left(g-\lambda _{\sigma}\right)-{\phi }^2 \left(g-\lambda _{\phi }\right)\right) \nonumber \\
&&+\frac{1}{4} \left({\sigma} ^4\left(g-\lambda _{\sigma }\right)^2 +{\phi} ^4 \left(g-\lambda _{\phi }\right){}^2+2 {\phi} ^2 {\sigma} ^2 \left(g^2+\left(\lambda _{\sigma }+\lambda _{\phi}\right) g-\lambda _{\sigma } \lambda _{\phi }\right)\right)\bigg]^{1/2}.\nonumber\\
\ea
One notices that the effective action (\ref{main1}), (\ref{main2}) has a very simple structure. As for the classical action, it is completely symmetric in $\phi$ and $\sigma$, and in particular the second and third lines of (\ref{main2}) are the same with $\phi\leftrightarrow\sigma$, as are the next two. 

We have specialised to FRW space-time, defined in terms of the scale factor $a(t)$,
\ba
\label{eq:metric}
ds^2 = -dt^2+a^2(t)d{\bf x}^2,
\ea
where we can define the Hubble rate $H=\dot{a}/a$ and the higher order gravitational tensors reduce to 
\ba
C^2 =0, \qquad G= 24 \left(\frac{\dot{a}}{a}\right)^2\frac{\ddot{a}}{a},\qquad R = 6\left[\left(\frac{\dot{a}}{a}\right)^2+\frac{\ddot{a}}{a}\right].
\ea
As explained in section \ref{sec:truncation}, partial integration of $\Box\times field$ type terms multiplying the logarithms in (\ref{main2}) reveals that they are beyond our level of approximation\footnote{This is true in unbounded space. Otherwise, there may be additional boundary terms.}. Also, the $\Box\times field$ terms that would appear in (\ref{main1}) integrate to zero. These have all been discarded already along with $C^2$. For the full general results, see Eq. (\ref{eq:fullres1}) and (\ref{eq:fullres2}).

Gravitational corrections enter through the second order operators $R\sim \frac{\dot{a}^2}{a^2}$, and the fourth order operators $G$, $C^2$, $R^2$, $\Box R$ $\sim \frac{\dot{a}^4}{a^4}$. We have truncated keeping this order, which we refer to as $\mathcal{O}(R^2)$ (see section \ref{sec:computation}). Ignoring these, the expression reduces to a Coleman-Weinberg-type effective action. In particular when also setting $g=0$, it is easy to see that we get the sum of two decoupled Coleman-Weinberg potentials for $\sigma$ and $\phi$.  
The quantum correction $\mathcal{L}_{eff}^{(1)}$ is suppressed by the phase space factor $1/(64\pi^2)$ and is in this sense small. Also reinstating physical units, $\hbar$ would appear in front of $\mathcal{L}_{eff}^{(1)}$. 

\subsection{A specific case: FRW, minimal coupling to gravity and no self-interaction}
\label{sec:specific}

The above expression in all generality may be applied to a number of cosmological scenarios and phenomena, but in the following we will simplify things somewhat by taking the values 
\ba
\xi_\sigma=\xi_\phi=1,\qquad\beta=\epsilon_1=\epsilon_2=\lambda_\sigma=\lambda_\phi=0,
\ea
and use the symmetry of the potential to solve for one of the fields in its equilibrium
\ba
\phi=0,
\ea
leaving the other field $\sigma$ to evolve freely. We emphasize that this is simply an appropriate parameter choice for the particular application we have in mind here, and not a further approximation. This allows us to write
\ba
\label{unbar}
\mathcal{L}_{eff}&=&-\f{\partial_\mu\sigma\partial^\mu\sigma}{2}-\f{m_\sigma^2}{2}\sigma^2+\Lambda+\alpha R\nonumber \\
&&+\f{1}{64\pi^2}\bigg\{\f{1}{24}\big(R-3 g\sigma^2\big)\big(R-3g\sigma^2-4m_\phi^2\big)\nonumber \\
&&+\bigg[-\bigg(m^2_\phi-\f{R}{6}+\f{g\sigma^2}{2}\bigg)^2+\f{G}{180}\bigg]\log\bigg(\f{m_\phi^2-\f{R}{6}+\f{g\sigma^2}{2}}{m^2_\phi}\bigg)
\nonumber \\
&&+\f{1}{24}\big(R-4m^2_\sigma\big)R+\bigg[-\bigg(m^2_\sigma-\f{R}{6}\bigg)^2+\f{G}{180}\bigg]\log\bigg(\f{m_\sigma^2-\f{R}{6}}{m^2_\sigma}\bigg)\bigg\}.
\ea
In fact, the entire last line can be discarded since an expansion in $R/M^2_{\pm}$ reveals that it is also $\mathcal{O}(R^3)$, with the understanding that the logarithm must be finite, i.e. $R/(6m_\sigma^2)<1$. Similarly from the first logarithm, and anticipating that $\sigma$ is to oscillate through zero, we must also demand $(R/6m_\phi^2)<1$. We will see in section \ref{sec:computation} that this is partly a result of our choice of renormalization scale.  We also mention here that for the derivative expansion to be reliable, one must require $g\lesssim 1$.

\section{Quantum corrected field dynamics}
\label{sec:dynamics}

\subsection{Dynamics of spectator fields}
\label{sec:spectator}

\begin{figure}
\begin{center}
\includegraphics[width=0.75\textwidth]{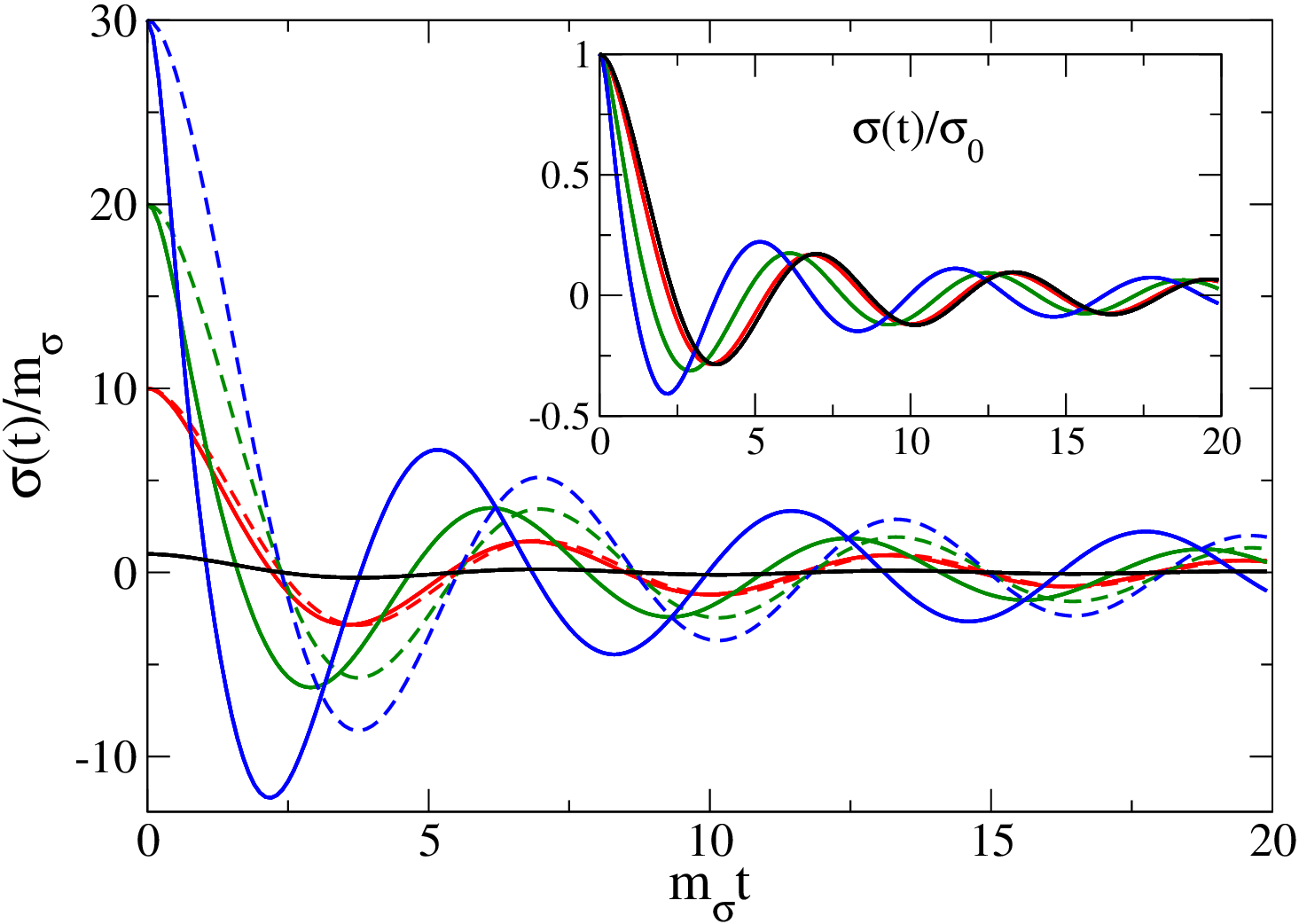}
\caption{The evolution of the spectator field $\sigma(t)$ for different values of $\sigma_0$ and different approximations, in a matter dominated Universe. We use $m_\phi/m_\sigma=2$, $g=1$, and $H_0/m_\sigma=\sqrt{1/2}$. The inset shows the same thing but rescaled by the initial values $\sigma_0$.}
\label{fig:Mat}
\end{center}
\end{figure}

We start by considering the case where the fields $\sigma$ and $\phi$ do not dominate the energy density. The obvious example of this is the subdominant curvaton during inflation. We assume that there is some other energy component determining the evolution of the background metric, in FRW the scale factor $a(t)$. We will consider three typical cases, radiation dominated, matter dominated and de Sitter space, where the scale factor evolves as
\ba
a_{\rm rad}(t) = a_0\left(\frac{t+t_0}{t_0}\right)^{1/2},\qquad a_{\rm mat}(t)= a_0\left(\frac{t+t_0}{t_0}\right)^{2/3},\qquad a_{\rm dS}(t)= a_0e^{H_0 t},
\ea
and we have the Hubble rate
\ba
H_{\rm rad} = \frac{1}{2(t+t_0)},\qquad H_{\rm mat}=\frac{2}{3(t+t_0)},\qquad H_{\rm dS}=H_0,
\ea 
and the higher order operators
\ba
R_{\rm rad}=0,\qquad R_{\rm mat}=\frac{4}{3(t+t_0)^2}\qquad R_{\rm dS}=12H_0^2,
\ea
and
\ba
G_{\rm rad}=-\frac{3}{2(t+t_0)^4},\qquad G_{\rm mat}=-\frac{64}{27(t+t_0)^4},\qquad G_{\rm dS}=24H_0^4.
\ea
Since the metric is given and we have initialized $\phi$ in its minimum, $\phi=0$, we only have to consider the equation of motion for $\sigma$ which reads
\ba
\label{bar1}
\ddot{\sigma}+3H\dot{\sigma}+m^2_\sigma\sigma&=&\f{1}{64\pi^2}\bigg\{\f{g\sigma}{2}\big(2m^2_\phi-R+3g\sigma^2\big)+g\sigma\f{\big(\f{G}{180}-\big(m^2_\phi-\f{R}{6}+\f{g\sigma^2}{2}\big)^2\big)}{m^2_\phi-\f{R}{6}+\f{g\sigma^2}{2}}\nonumber \\
&&-2g\sigma\big(m^2_\phi-\f{R}{6}+\f{g\sigma^2}{2}\big)\log\bigg(\f{m^2_\phi-\f{R}{6}+\f{g\sigma^2}{2}}{m^2_\phi}\bigg)\bigg\}=0,
\ea
At tree level this reduces to
\ba
\label{treelevel}
\ddot{\sigma}+3H\dot{\sigma}+m^2_\sigma\sigma&=&0.
\ea
and there is then no dependence on $g$ or $m_\phi$, and $\sigma$ does not feel the presence of the $\phi$, to which it is coupled. 

\begin{figure}
\begin{center}
\includegraphics[width=0.75\textwidth]{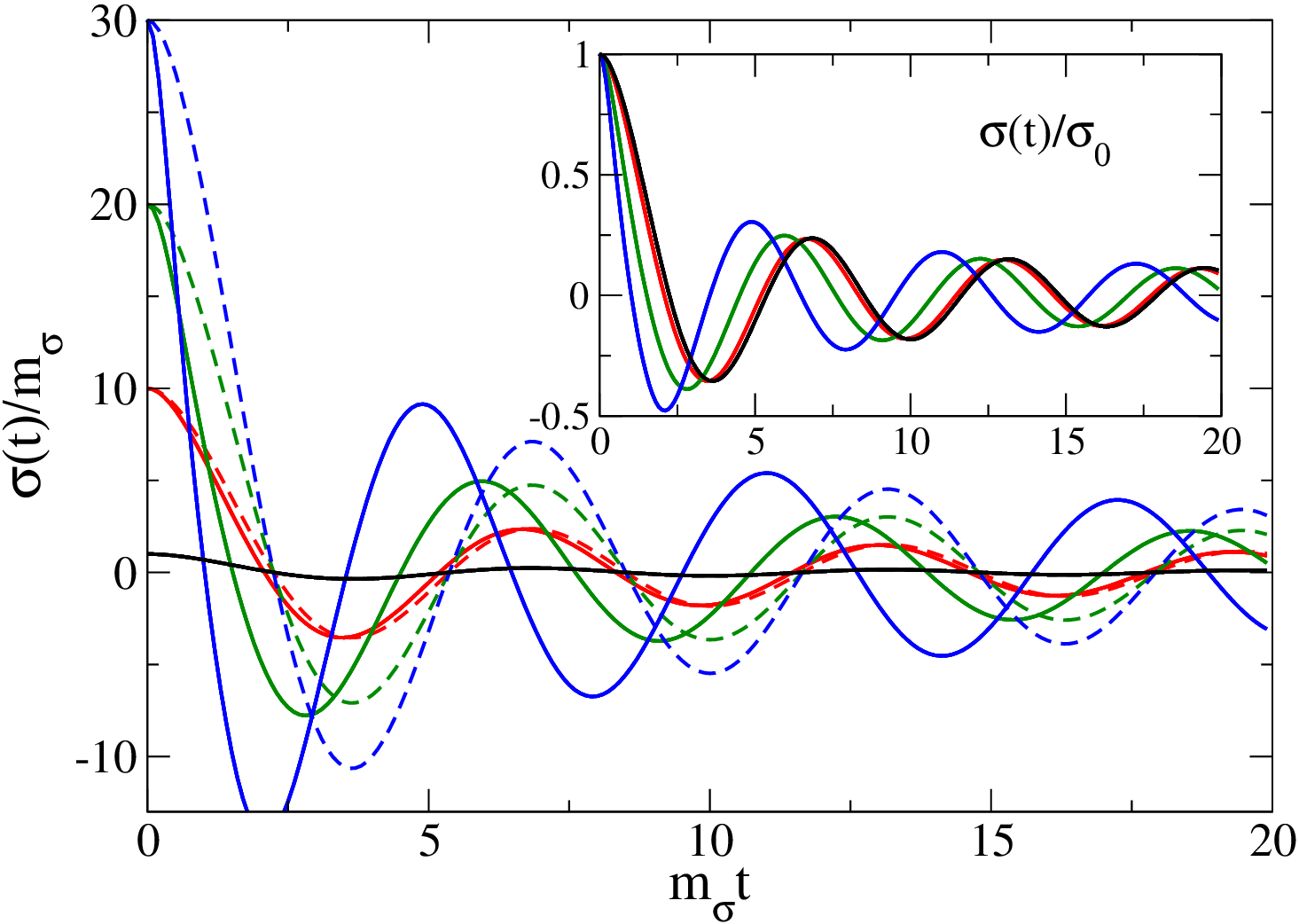}
\caption{The evolution of the spectator field $\sigma(t)$, for different values of $\sigma_0$ and different approximations, in a radiation dominated Universe. We use $m_\phi/m_\sigma=2$, $g=1$, and $H_0/m_\sigma=\sqrt{1/2}$. The inset shows the same thing but rescaled by the initial values $\sigma_0$.}
\label{fig:Rad}
\end{center}
\end{figure}

We solve the evolution equations in the three cosmological backgrounds, identifying four levels of approximation: Tree level is (\ref{treelevel}), order $H^0$ ignores gravitational operators in the quantum corrections, order $H^2$ includes all occurrences of $R$ and finally order $H^4$ also includes $G$. We choose the parameters $m_\phi/m_\sigma=2$, $H_0/m_\sigma=\sqrt{1/2}$, $g=1$. $m_\sigma$ sets the scale, and $H_0$ is chosen so that the argument of the logarithm is always positive. In particular $R_0/(6m_\phi^2)=1/2$.

We then vary the initial amplitude $\sigma_0/m_\sigma$, and since (\ref{treelevel}) is linear the evolution is simply rescaled. The full dynamics is no longer linear, and such a rescaling no longer works. 

\begin{figure}
\begin{center}
\includegraphics[width=0.75\textwidth]{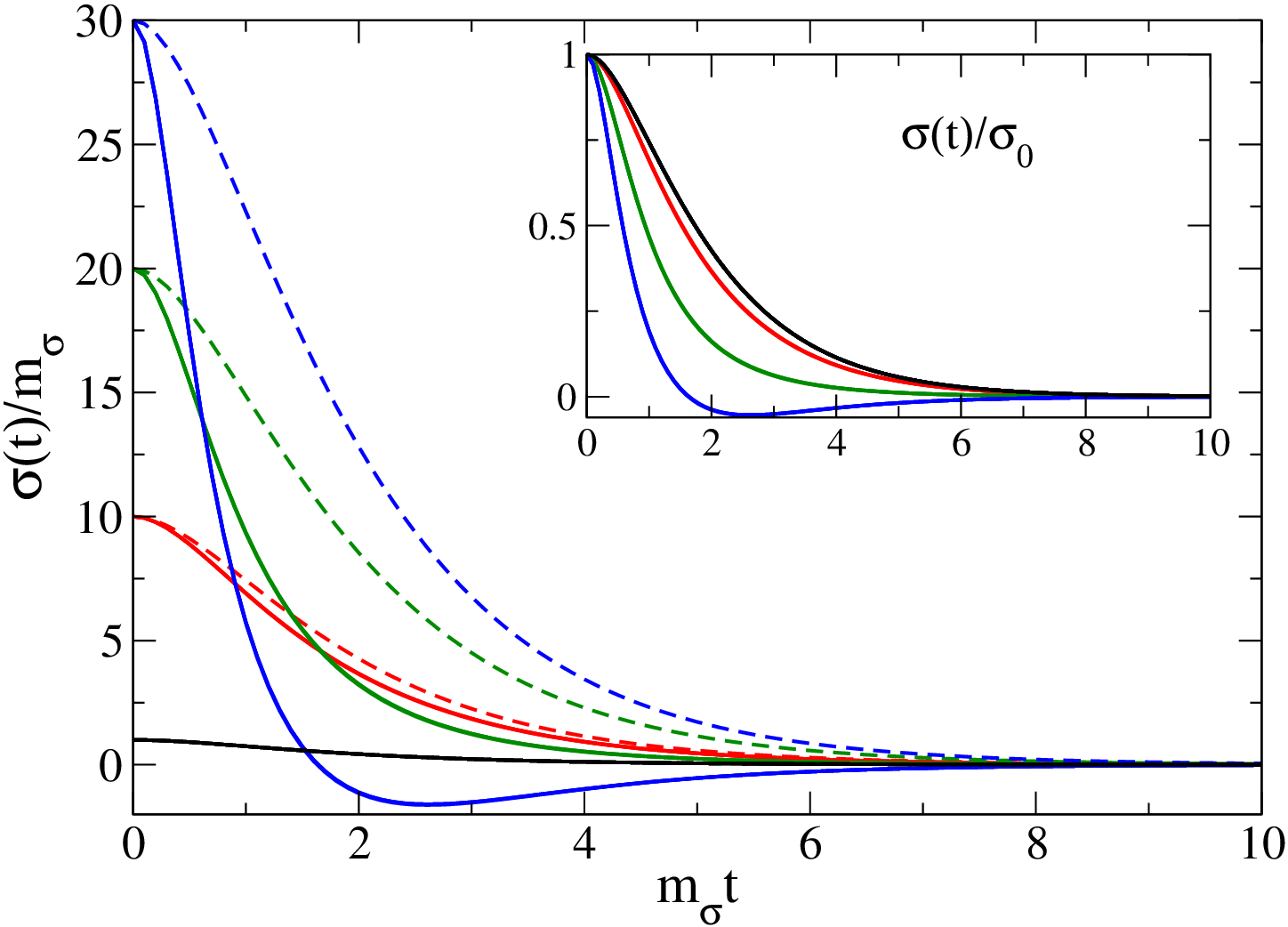}
\caption{The evolution of the curvaton field $\sigma(t)$, for different values of $\sigma_0$ and different approximations, in a de Sitter Universe. We use $m_\phi/m_\sigma=2$, $g=1$, and $H_0/m_\sigma=\sqrt{1/2}$. The inset shows the same thing but rescaled by the initial values $\sigma_0$.}
\label{fig:Inf}
\end{center}
\end{figure}

In Fig.~\ref{fig:Mat}, we show the evolution for a matter dominated background, in Fig.~\ref{fig:Rad} for radiation and in Fig.~\ref{fig:Inf} in de Sitter space. We see that the field performs a damped oscillation, which in the de Sitter case is even over-damped. Choosing $\sigma_0/m_\sigma =1, 10, 20, 30 $, we get the black red, green and blue curves, respectively and where the dashed line signifies the classical tree level result and the full lines the quantum corrected ones. In the insets, we show the same curves rescaled with $\sigma_0$, to demonstrate the deviation from the linear evolution. 

The quantum corrections are numerically relatively small at these values of parameters, which is simply due to the overall factor $1/(64\pi^2)$. We also find that different approximations of order $H^{0,2,4}$ as defined above are essentially identical (they all fit within the full lines). This means that the quantum effects are due to the non-gravitational corrections, i.e. the Minkowski space effective potential. 

Our choice of renormalization conditions (see section \ref{sec:renormalization}) tunes the counterterms so that the effective potential matches the tree level one close to $\sigma=0$. But for field values far from zero, the effective potential is much steeper and non-quadratic, leading to non-linear oscillation. These non-linearities are a quantum effect. 

Choosing a different renormalization point changes the effective potential to match at some other value of $\sigma$, but then that effective potential has large corrections near $\sigma=0$. The different choices are physically distinct, and should be based on what the value of the physical, renormalized parameters are taken to be.  

This concludes our study of spectator fields in a given gravitational background. We now turn to the case where $\sigma$ and $\phi$ dominate the energy density.

\subsection{Quantum corrected Friedmann equations}
\label{sec:friedmann}

The quantum effects in the effective action also lead to corrections to the Friedmann equations, which follow from taking the variation\footnote{Note that one must vary w.r.t. the general metric $g^{\mu\nu}$ and only afterwards specialize to FRW.}
\ba
\frac{\delta \Gamma[g^{\mu\nu},\phi,\sigma]}{\delta g^{\mu\nu}}=0.
\ea
Since the effective action has curvature terms up to $\mathcal{O}(R^2)$ the Friedmann equation is now fourth order in time derivatives (see \ref{sec:Friedmann}). One may reduce the fourth order Friedmann equation to a second order one keeping all the physical solutions \cite{fourdiv}, but the procedure is rather involved. Here we settle for $\mathcal{O}(R)$. We will also restrict ourselves to the simplified case (\ref{unbar}) for the field dynamics, but an analogous procedure of course applies to the general result (\ref{main1}) and (\ref{main2}). 
We find from the ``00'' component
\ba
\label{ein1}
&&\f{\dot{a}^2}{a^2}\bigg\{1-\f{G_N}{12\pi}\bigg[\f{g\sigma^2}{2}-\bigg(m^2_\phi+\f{g\sigma^2}{2}\bigg)\log\bigg(1+\f{g\sigma^2}{2m^2_\phi}\bigg)\bigg]\bigg\}+\f{\dot{a}}{a}\bigg\{\f{G_N}{12\pi}\frac{\dot{\sigma}}{\sigma}g\sigma^2\log\bigg(1+\f{g\sigma^2}{2m^2_\phi}\bigg)\bigg\}\nonumber \\&&=\f{\Lambda'}{3}+\f{8\pi G_N}{3}\bigg(\f{\dot{\sigma}^2}{2}+\f{m^2_\sigma}{2}\sigma^2 \bigg )\nonumber \\&&\qquad\qquad\qquad\qquad-\f{G_N}{12\pi}\bigg[\f{g\sigma^2\big(4m^2_\phi+3g\sigma^2\big)}{16}-\f{1}{2}\big(m^2_\phi+\f{g\sigma^2}{2}\big)^2\log\bigg(1+\f{g\sigma^2}{2m^2_\phi}\bigg)\bigg].
\ea
This result has a straightforward interpretation. When neglecting the loop contribution (terms proportional to $G_N/(12\pi)$), the equation reduces to the well-known classical Friedmann equation
\ba
\label{clasfriedmann1}
\f{\dot{a}^2}{a^2}=\f{8\pi G_N}{3}\bigg(\f{\dot{\sigma}^2}{2}+\f{m^2_\sigma}{2}\sigma^2\bigg )+\f{\Lambda'}{3}=\f{8\pi G_N}{3}\rho+\f{\Lambda'}{3},
\ea
where $\rho=T_{00}$ is the classical energy density in the fields. The quantum corrections involve quite non-trivial field-dependence and in addition have explicit dependence on $H=\dot{a}/a$. We note that occurrences of $R^2$ (for instance coming from expanding the logarithms) have been neglected through neglecting derivative orders beyond two.

Similarly, for the "$ii$ " component we find, after using (\ref{ein1}) to simplify the expression
\ba
\label{ein2}
&&\f{\ddot{a}}{a}\bigg\{1+\f{ G_N}{24\pi }\bigg[\left(2m^2_\phi+g\sigma^2\right)\log\bigg(1+\f{g\sigma^2}{2m^2_\phi}\bigg)-g\sigma^2\bigg]\bigg\}
+\f{\dot{a}}{a}\bigg\{\f{G_N}{24\pi}\frac{\dot{\sigma}}{\sigma}g\sigma^2\log\bigg(1+\f{g\sigma^2}{2m^2_\phi}\bigg)\bigg\}\nonumber\\
&&=\f{\Lambda'}{3}+\f{4 \pi G_N}{3}\big(-2\dot{\sigma}^2+m^2_\sigma\sigma^2\big)\nonumber \\
&&
+\f{G_N}{24\pi}\bigg[-\f{g\sigma^2\big(4m^2_\phi+3g\sigma^2\big)}{8}-\f{g^2\sigma^2\dot{\sigma}^2}{m^2_\phi+\f{g\sigma^2}{2}}
+\bigg(\bigg(m^2_\phi+\f{g\sigma^2}{2}\bigg)^2-\big(g\dot{\sigma}^2+g\ddot{\sigma}\sigma\big)\bigg)\log\bigg(1+\f{g\sigma^2}{2m^2_\phi}\bigg)\bigg],\nonumber\\
\ea
Again, the interpretation is clear, and neglecting the quantum corrections, we find the standard expression
\ba
\label{clasfriedmann2}
\f{\ddot{a}}{a}=\f{4\pi G_N}{3}\left(-2\dot{\sigma}^2+m^2_\sigma\sigma^2\right)+\f{\Lambda'}{3}=-\f{4\pi G_N}{3}(\rho+3p)+\f{\Lambda'}{3},
\ea
with the classical pressure $p=T_{ii}/a^2$ (no summation). The expression now also involves second derivatives of the field. There is again a term proportional to $\dot{a}/a$, which vanishes in the classical limit.

If we choose to separate the effective lagrangian (\ref{unbar}) in the form $S=\alpha R+\mathcal{L}^{\rm rest}(\sigma, R)$, Eqs.~(\ref{ein1}) and (\ref{ein2}) correspond to the classical Friedmann equations, where we replace the right-hand side by the appropriate components of the quantum corrected energy-momentum tensor. The equation of motion from the renormalized Lagrangian then gives 
\begin{gather}\label{eq:Etensor}
\f{\delta S}{\delta g^{\mu\nu}}=0\nonumber \\ \Leftrightarrow\quad
2\alpha G_{\mu\nu}=-\frac{2}{\sqrt{-g}}\frac{\delta \int d^4x\,\sqrt{-g} \mathcal{L}^{\rm rest}(\sigma,R)}{\delta g^{\mu\nu}}=\langle T_{\mu\nu}\rangle.
\end{gather}
Since this equation is renormalized and thus perfectly finite, the contracted Bianchi identity guarantees covariant conservation of the energy-momentum. However, if one starts from the unrenormalized Einstein equation
\begin{equation}
2\alpha_0 G_{\mu\nu}=\langle T_{\mu\nu}\rangle_0
\end{equation}
A counter term $\delta T_{\mu\nu}$ must be introduced on the right-hand side to cancel the divergences, which might break covariant conservation (or produce other unwanted artefacts). The latter approach is used for example in \cite{Baacke:2010bm,MolinaParis:2000zz} where energy was indeed found to be conserved, but this had to be checked and was not trivially deducible from the formalism.

\subsection{Self-consistent dynamics of field and scale factor}
\label{sec:energy}

In the following, we will solve the equation of motion for $\sigma$ together with (\ref{ein2}) for the tree level case (\ref{treelevel}) and the quantum corrected equation (\ref{bar1}). We take the parameters $M_{\rm pl}/m_\sigma=100$, $g=1$, $m_\phi/m_\sigma=2$ and vary $\sigma_0/m_\sigma$. We define the reduced Planck mass as $M_{\rm pl}^2 = 1/(8\pi^2 G_N)$. The initial value of the Hubble rate $H_0$ is given by (\ref{ein1}), and we simply set $\dot{\sigma}_0=0$.

\begin{figure}
\begin{center}
\includegraphics[width=0.45\textwidth]{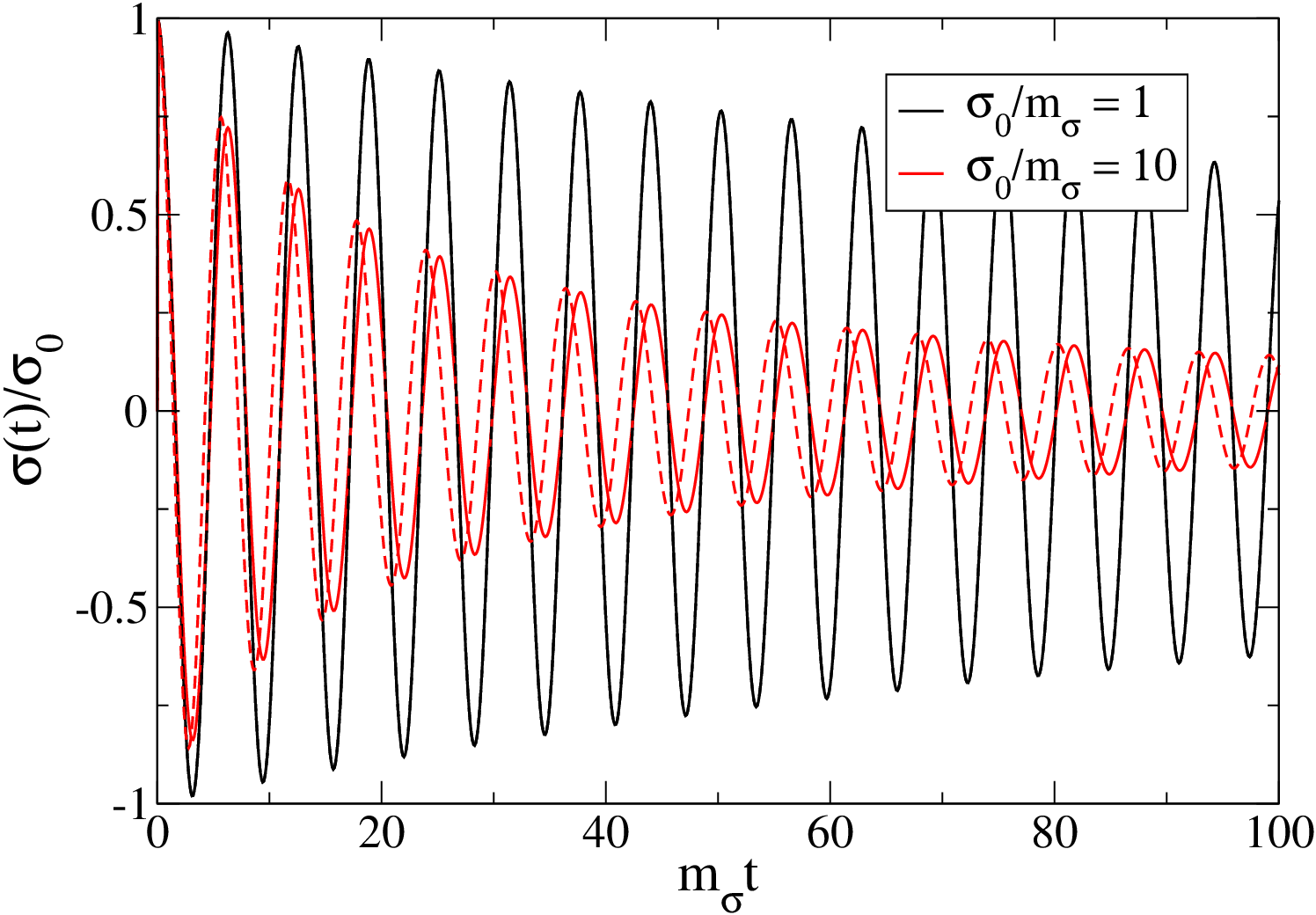}
\includegraphics[width=0.45\textwidth]{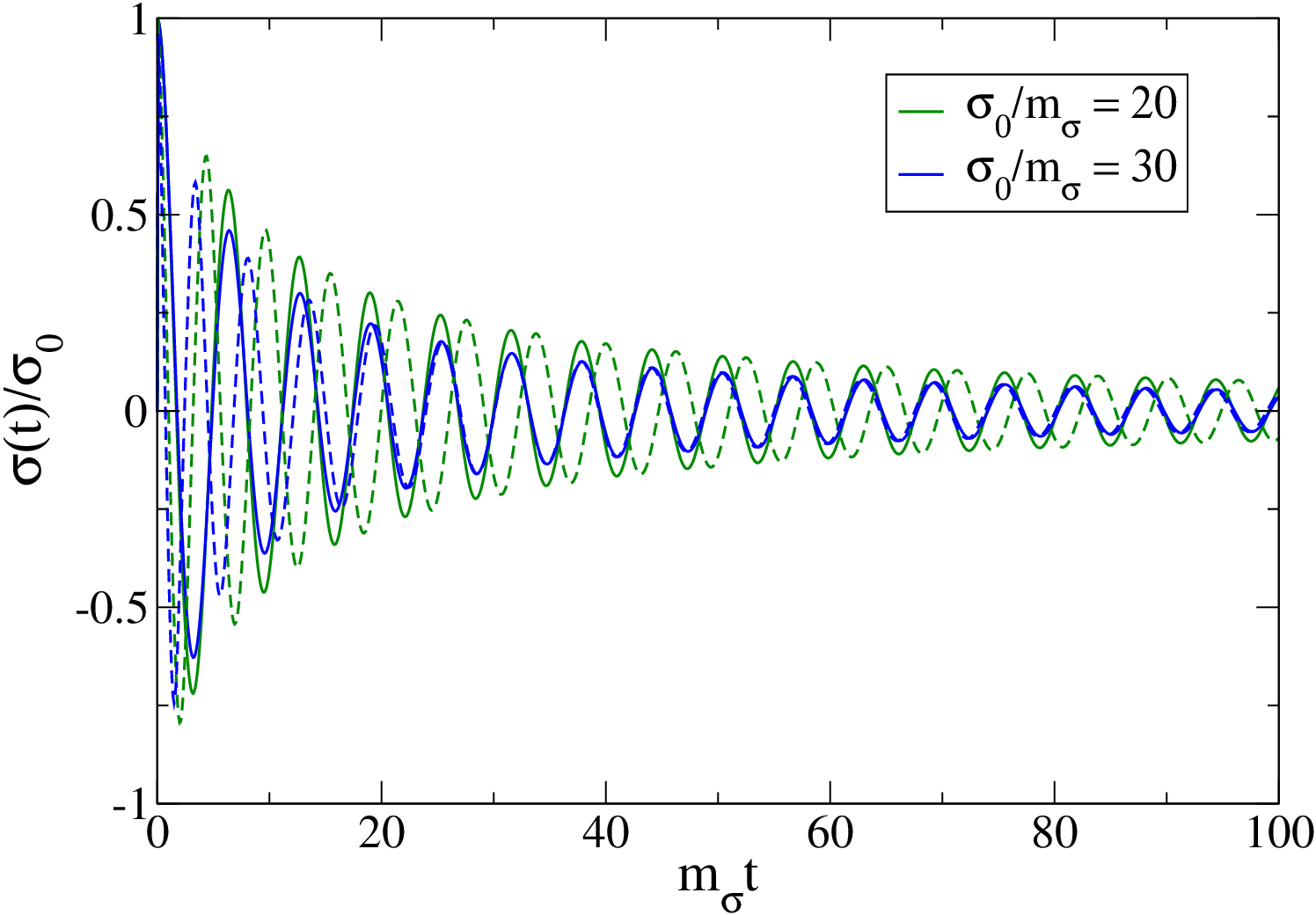}
\caption{The evolution of the inflation field $\sigma(t)$ rescaled by its initial value $\sigma_0$, for different values of $\sigma_0$ and different approximations, when the scale factor is given by the tree level (full lines) and quantum corrected (dashed) Friedmann equation. We use $m_\phi/m_\sigma=2$, $g=1$.}
\label{fig:sig}
\end{center}
\end{figure}

Fig.~\ref{fig:sig} shows the field $\sigma$ using different initial amplitudes $\sigma_0$, and rescaled by this amplitude. Whereas for the spectator field above this rescaling is exact for the tree level equation, this is no longer the case here, since the Friedmann equation has a non-linear dependence on the initial amplitude. In the left-hand plot we show $\sigma_0/m_\sigma=1$ and $10$, and only for the larger amplitude is the quantum corrected evolution (dashed line) discernably different from the tree level evolution (full line). In the right-hand plot we see that this trend persists for even larger initial field amplitude.

\begin{figure}
\begin{center}
\includegraphics[width=0.45\textwidth]{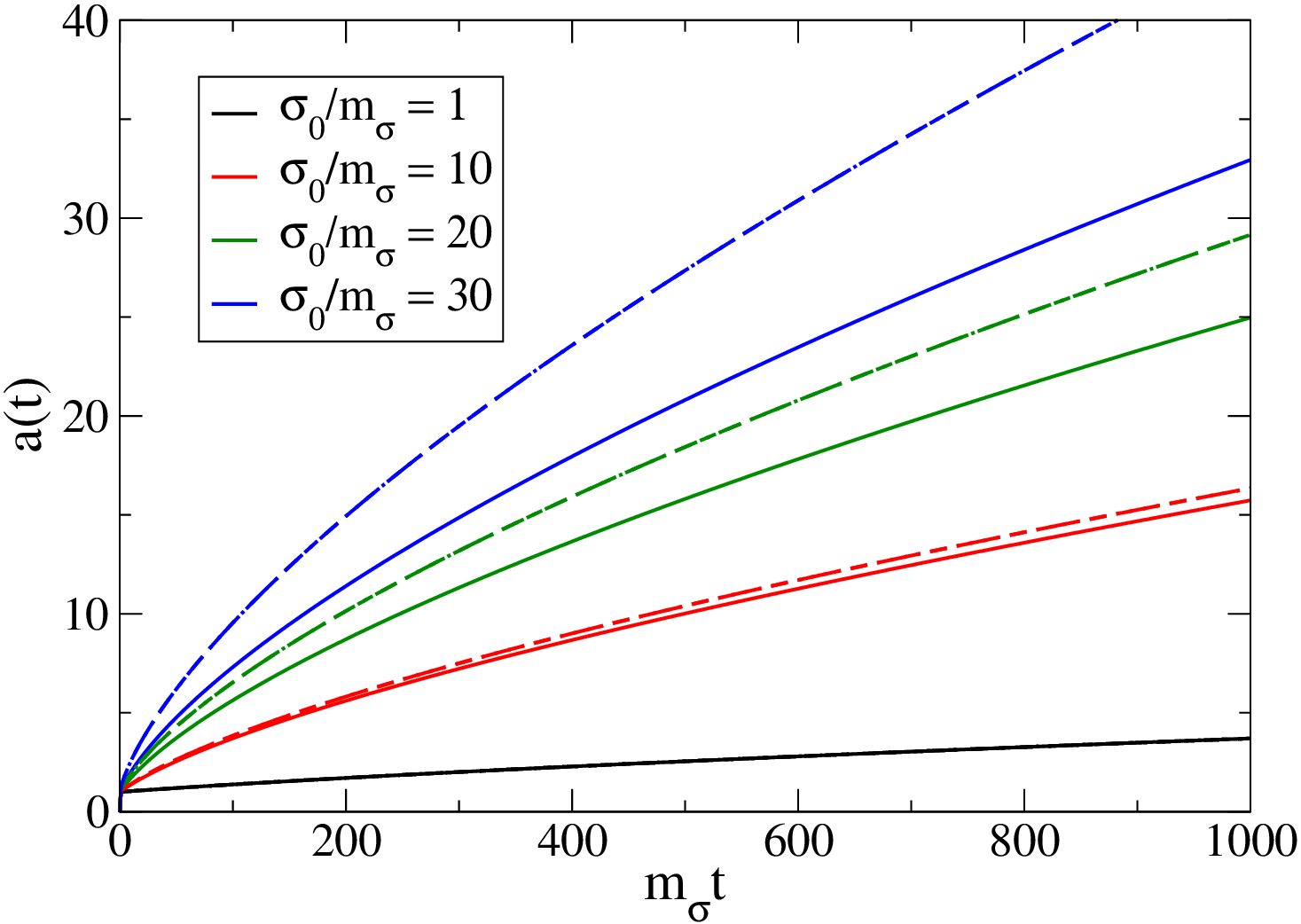}
\caption{The evolution of the scale factor $a(t)$ from the tree level (full lines) and quantum corrected (dashed) approximations.}
\label{fig:a}
\end{center}
\end{figure}

In Fig.~\ref{fig:a} we show the corresponding scale factor $a(t)$, for all the different initial amplitudes and for the tree level (full) and quantum corrected (dashed) equations. All curves turn out to evolve as matter domination $\propto t^{2/3}$, but with different rates. Again the quantum correction is negligible for the smallest amplitude (black lines), whereas there is a significant effect at larger amplitudes.

\subsection{Inflation}
\label{sec:inflation}

\begin{figure}
\begin{center}
\includegraphics[width=0.45\textwidth]{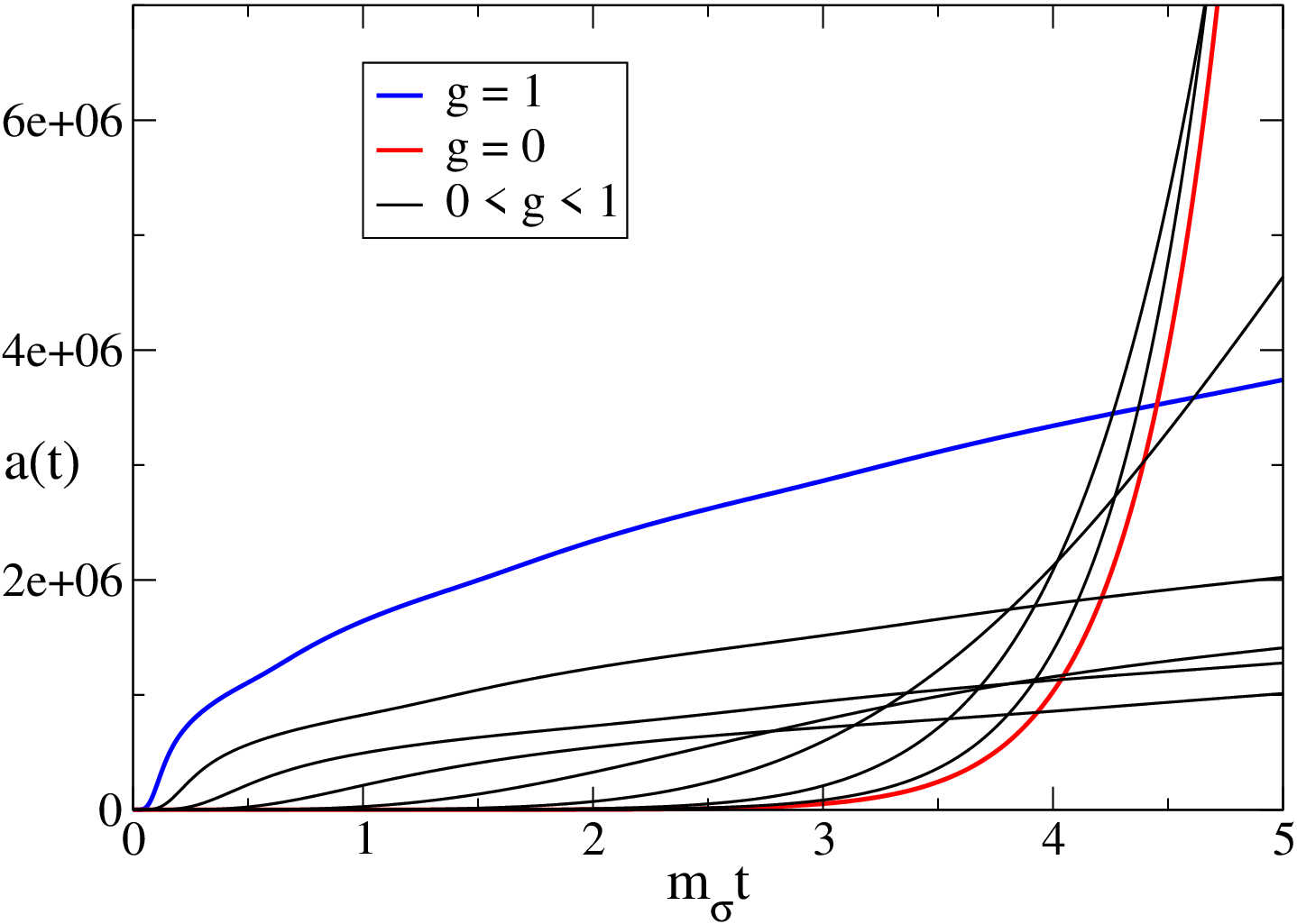} $\quad$
\includegraphics[width=0.45\textwidth]{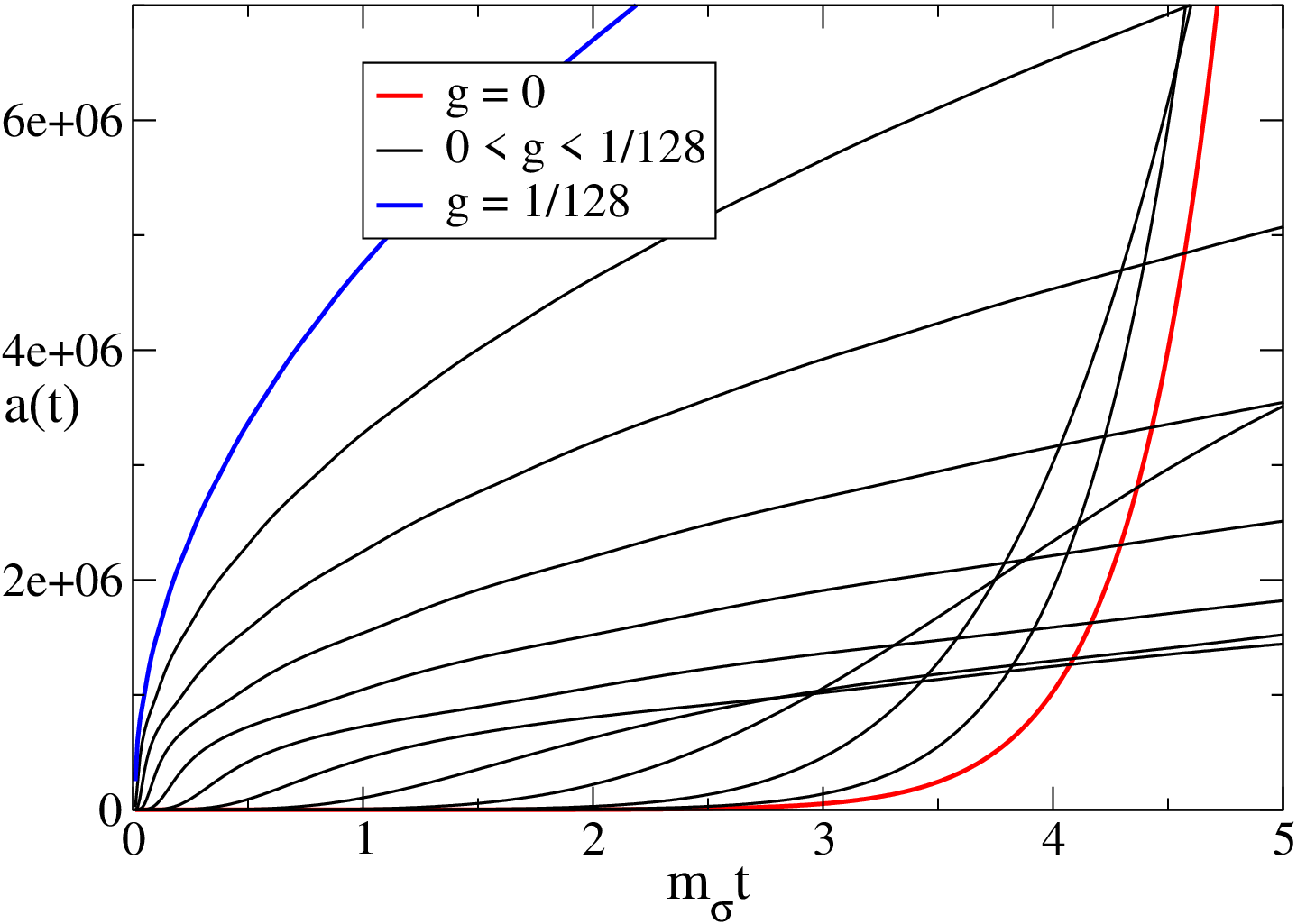}
\caption{The evolution of the scale factor when $\sigma_0=10 M_{\rm pl}$,
for different values of $g$. Left, for $M_{\rm pl}/m_\sigma=100$. Right,
for $M_{\rm pl}/m_\sigma=10^5$. }
\label{fig:inflation}
\end{center}
\end{figure}

Of particular importance for cosmology is the possibility of inflation at
the very earliest stages of the evolution of the Universe. In a simple
slow-roll computation, we have for the tree level quadratic potential
\ba
V=\frac{1}{2}m_\sigma^2\sigma^2,\qquad \qquad \epsilon = \frac{2M_{\rm
pl}^2}{\sigma^2}.
\ea
Hence, neglecting quantum corrections, inflation will arise for
$\sigma_0>\sqrt{2}M_{\rm pl}$, when $\epsilon<1$. In
Fig.~\ref{fig:inflation} we show the evolution of $a(t)$ for
$\sigma_0=10M_{\rm pl}$ for early times, for $M_{\rm pl}/m_\sigma=100$
(left) and $10^5$ (right). Indeed in the tree-level case (or
non-interacting, $g=0$, red line), inflation is prolonged and leads to
exponential growth of the Universe. Now we gradually add the interactions
by increasing $g$ (full, black lines). We see that inflation becomes
weaker and weaker, and finally disappears for $g\simeq 1$ (left) and
$g\simeq 10^{-2}$ (right). Hence because of quantum effects, interactions
with another field $\phi$ may prevent inflation from taking place, at
least in this case for $\sigma_0<10M_{\rm pl}$. This is a well-known
result, which has now also been confirmed in the present framework.

The value $M_{\rm pl}/m_\sigma\simeq 10^5$ is what is required for the
spectrum of the CMB to be correctly reproduced by quadratic inflation (see
for instance \cite{Liddle:2000cg}). If the inflaton is coupled to other
fields, this conclusion must be re-evaluated in the context of quantum
corrections to the potential. Since an analytic solution of the slow-roll
field evolution from the end of inflation back to the era of horizon
crossing is not readily available, we will postpone a detailed analysis of
this point to future work.

\subsection{Field fluctuations from the propagator}
\label{sec:propagator}

A crucial observable for inflationary physics is the spectrum of density perturbations, which arise from the quantum fluctuations of the inflaton or a spectator field like the curvaton. From the effective action point of view these are encoded in the field propagator (or two-point function), which follows from taking the second variation of $\Gamma$ w.r.t. the fields, i.e.
\ba
\mathcal{G}_{\chi_1\chi_2}^{-1}=\frac{\delta^2 \Gamma[g^{\mu\nu},\phi,\sigma]}{\delta\chi_1\delta\chi_2},
\ea
where $\chi_{1,2}$ can be $\phi$ or $\sigma$. This gives a matrix equation for the propagator
\ba
\mathcal{G}^{-1}(x,x')\mathcal{G}(x',x'')=
\left(\begin{array}{cc}
\frac{\delta^2 \Gamma}{\delta\sigma\delta\sigma}&
\frac{\delta^2 \Gamma}{\delta\sigma\delta\phi}\\
\frac{\delta^2 \Gamma}{\delta\phi\delta\sigma}&
\frac{\delta^2\Gamma}{\delta\phi\delta\phi}
\end{array}
\right)(x,x')
\left(\begin{array}{cc}
G_{\sigma\sigma}&
G_{\sigma\phi}\\
G_{\phi\sigma}&
G_{\phi\phi}
\end{array}\right)(x',x'')= \frac{\delta(x-x'')}{\sqrt{-g}},
\ea
where an integral over the $x'$ indexes is left implicit. At the one-loop level employed here, the propagator equation is equivalent to solving the equation of motion for each field mode $f_k(t)$ of the mass-eigenstate fields
\ba
\Phi_\pm(x,t)=\Phi_\pm(t)+\int \frac{d^3k}{(2\pi)^3}\left[a_k^\pm f_k^\pm(t)+(a_k^\pm)^\dagger (f_k^\pm)^*(t)\right],
\ea
where $a_k^\pm$ are creation operators and the equation of motion is
\ba
-\ddot{f}_k-3H\dot{f}_k-\frac{k^2}{a^2}f_k-M_\sigma^2(\sigma,\phi)f_k=0.
\ea
for some $M_\sigma^2(\sigma,\phi)$. For sufficiently large momenta $k$ the mass term can be neglected as is usually done, but at horizon crossing $H=k$ we have $H>M(\sigma,\phi)$, and so formally the truncation leading to the effective action in (\ref{main1}, \ref{main2}) no longer holds.  But since the equation for the propagator in that limit is the free field equation used in standard computations of the spectrum we just conclude that we have no quantum corrections for this observable, other than what enters via the altered evolution of $a(t)$ and $\sigma(t)$.

For the 1PI effective action, one first solves for the mean field, and this is then put into the propagator equation which is then solved. At higher loop order, additional interactions will make the propagator equation non-trivial through a non-linear dependence on the mean field.

\section{Computation of the effective action and equations of motion}
\label{sec:computation}

\subsection{Effective action for $\varphi^4$ theory}
\label{sec:eff1scalar}

We will first discuss the relevant points in deriving the curved space effective action for $\varphi^4$ theory to one-loop order. Our discussion follows closely that presented in \cite{Kirsten:1993jn}. For more details, see also \cite{ParkerToms,Hu:1984js,Elizalde:1994ds}. We start from an action with a matter part
\begin{equation}
S_m[\varphi,g^{\mu\nu}]= \int d^4x\sqrt{-g}~\bigg[-\frac{1}{2}g^{\mu\nu}\partial_\mu\varphi\partial_\nu\varphi+\eta\Box\varphi^2-\frac{m^2}{2}\varphi^2-\f{1}{2}(1-\xi)\f{R}{6}\varphi^2-\f{\lambda}{4!}\varphi^4\bigg]
\end{equation}
and a gravitational part
\begin{equation}
S_g[g^{\mu\nu}]\equiv \int d^4x\sqrt{-g}~\bigg[\Lambda+\alpha R+\beta R^2+\epsilon_1 C^2+\epsilon_2 G+\kappa \Box R\bigg].
\end{equation}
Our aim is to write the effective action as a loop expansion,
\begin{equation}
\Gamma[\varphi,g^{\mu\nu}]=\int d^4 x \sqrt{-g} \mathcal{L}_{eff}=\Gamma^{(0)}[\varphi,g^{\mu\nu}]+\Gamma^{(1)}[\varphi,g^{\mu\nu}]+\cdots,
\end{equation}
where the metric $g^{\mu\nu}$ is considered as a classical field. The standard method is to start from the generating functional for connected diagrams, use an expansion of the action to quadratic order, discard the one-particle tadpoles and finally perform a Legendre transformation to obtain the results \cite{Peskin:1995ev}
\begin{equation}
\label{eq:gammas}
\Gamma^{(0)}[\varphi,g^{\mu\nu}]=S_m[\varphi,g^{\mu\nu}]+ S_g[g^{\mu\nu}]+\delta S[\varphi,g^{\mu\nu}],\quad\Gamma^{(1)}[\varphi,g^{\mu\nu}]=\f{i}{2}\text{Tr log}~G^{-1}(x,x'),
\end{equation}
where $G(x,x')$ is the Feynman propagator satisfying
\begin{equation}
\bigg[-\Box+m^2+\frac{1}{6} R(1-\xi )+\lambda\f{\varphi^2}{2}\bigg]G(x,x')=\f{\delta(x-x')}{\sqrt{-g}}.
\end{equation}
One may choose different boundary conditions (i.e choose different vacuum states) for the propagator in (\ref{eq:gammas}) but this will turn out irrelevant for our approximation. The counter terms enter as,
\begin{align}
\delta S[\varphi,g^{\mu\nu}]=\int d^4x\sqrt{-g}~\bigg[&-\frac{\delta Z}{2}g^{\mu\nu}\partial_\mu\varphi\partial_\nu\varphi+\delta\eta\Box\varphi^2-\frac{\delta m^2}{2}\varphi^2+\delta\xi\f{R}{12}\varphi^2-\f{\delta\lambda}{4!}\varphi^4\nonumber \\&+\delta\Lambda+\delta\alpha R+\delta\beta R^2+\delta\epsilon_1 C^2+\delta\epsilon_2 G+\delta\kappa \Box R\bigg],
\end{align}
and are introduced for the purposes of renormalization. In order to find an expression for $\Gamma^{(1)}$, we use the heat kernel method introduced for curved spacetime in \cite{Toms:1982af} and write the trace of a logarithm in (\ref{eq:gammas}) as a proper-time integral over a yet undefined kernel function $K$
\begin{equation}\label{hkint}
\f{i}{2}\text{Tr log}~G^{-1}(x,x')=-\f{i}{2}\mu^{4-n}\int d^n x\sqrt{-g}\int_0^\infty \f{d\tau }{\tau}K(\tau;x,x).
\end{equation}
Because of the divergent behaviour that occurs in four dimensions for $\Gamma^{(1)}$, we have dimensionally regularized the above integral to the dimension $n=4-\epsilon$ and added an arbitrary scale $\mu$ in order to maintain the proper dimension of the action. In appendix \ref{sec:HK} we present the heat kernel method in full detail. Here we simply state the result which reads
\begin{equation}\label{HK}
K(\tau;x,x)=i\f{\Omega(\tau;x,x)e^{-iM^2\tau}}{(4\pi i\tau)^{n/2}},
\end{equation}
where $M$ is the effective mass
\begin{equation}
M^2=m^2 -\f{\xi}{6}R+\lambda\f{\varphi^2}{2},
\end{equation}
and the $\Omega$'s have a small proper-time expansion
\begin{equation}\label{hamidew}
\Omega(\tau;x,x)=\sum_{k=0}^\infty a_k(x,x)(\tau i)^k.
\end{equation}
Substantial literature exists on these coefficients \cite{Minakshisundaram:1949xg} and \cite{DeWitt:1965jb} and results for the first four can be found for example in \cite{Jack:1985mw}. At this point would like to stress the generality of the above procedure. As we elaborate in section \ref{sec:eff2scalar} the above procedure is not merely suited for real scalar fields, but can be equally well applied to a large class of operators, including fields of higher spin, gauge theories and quantum gravity. See \cite{Barvinsky:1985an} for a detailed account. Inserting the expansion (\ref{hamidew}) into (\ref{HK}) and performing the proper time integral in (\ref{hkint}) we get for the effective action
\begin{equation}\label{series expansion}
\Gamma^{(1)}[\varphi,g^{\mu\nu}]=\int d^nx \sqrt{-g}~\f{1}{2(4\pi)^{n/2}}\bigg(\f{M}{\mu}\bigg)^{n-4}\sum^\infty_{k=0}M^{4-2k}a_k(x,x)\Gamma(k-n/2).
\end{equation}
Since we made use of a small proper time expansion to approximate the entire integral over the heat kernel, some comments on the validity of the procedure are in order. As is evident from the divergences of the Gamma function on the right hand side of (\ref{series expansion}), the small $\tau$ part of the heat kernel probes only the local region of space-time, and may miss some global aspects. For example, as is argued in Appendix \ref{sec:HK} boundary conditions are not seen by the expansion (\ref{hamidew}). The result for the full heat kernel is in fact non-local and has been developed recently, \cite{Barvinsky:1993en} and references therein, and applied to quantum scalar fields in \cite{Dalvit:1994gf}. The ansatz (\ref{HK}) is, however, a valid approximation for the full effective action when the fields are slowly varying \cite{Barvinsky:1994ic,Guven:1986gi,Matyjasek:2009mx}\footnote{Fields are smooth on the scale of $M^2$.}. The coefficients\footnote{Not to be confused with creation operators $a_k$ for quantum fields. We have adopted established notational conventions.} $a_k$ consist of geometric tensors and derivatives of $\varphi^2$ and from (\ref{eq:hkcoeff}) can be read to be $a_0=1$, $a_1=0$ and
\begin{equation}\label{aa1}
a_2(x,x)=-\f{\lambda}{12}\Box \varphi^2+(5\xi+1)\f{\Box R}{180}+\f{3C^2-G}{360},
\end{equation}
where we have defined $C^2$ and $G$ in (\ref{eq:CG})\footnote{Even when we have analytically continued to $n$ dimensions, we continue to use the four dimensional definition for $C^2$.}. 
The approximation to be used in this paper is to include only the first three terms of the right hand side of (\ref{series expansion}) in $\Gamma^{(1)}$, i.e. $a_0$, $a_1$ and $a_2$, neglecting $a_3$ and beyond, and then subsequently neglecting terms in $a_2$ that also appear in $a_3$ (see section \ref{sec:truncation}).
This gives us the effective action for a $\varphi^4$ theory to one-loop order
\begin{align}
\label{aa2}
\Gamma[\varphi,g^{\mu\nu}]&=S_m[\varphi,g^{\mu\nu}]+S_g[g^{\mu\nu}]+\delta S[\varphi,g^{\mu\nu}]\nonumber \\&+\int d^nx \sqrt{-g}~\f{1}{64\pi^2}\bigg\{-M^4\bigg[\log\bigg(\f{M^2}{\tilde{\mu}^2}\bigg)-\f{3}{2}\bigg]-2a_2(x,x)\log\bigg(\f{M^2}{\tilde{\mu}^2}\bigg)\bigg\},
\end{align}
where we defined a new arbitrary scale $\tilde{\mu}$
\begin{equation}\label{arbscale}
\log(\tilde{\mu}^2)=\f{2}{4-n}-\gamma+\log(4\pi\mu^2).
\end{equation}
From the right hand side of equation (\ref{aa2}) one can immediately see that the (regularized, but still unrenormalized) effective action consists of the well-known Coleman-Weinberg effective potential \cite{Coleman:1973jx} and corrections coming from field derivatives and gravity contributions. We reserve the closer investigation of the implications of our truncation in (\ref{aa2}) to section \ref{sec:truncation}. Renormalization will be discussed in section \ref{sec:renormalization}.

\subsection{Effective action for the two-scalar field model}
\label{sec:eff2scalar}

Before proceeding, we make some comments on the power and generality of the procedure. Following \cite{Steinwachs:2011zs} we can write the one loop effective action for a general theory with a multiplet of fields $\psi^A$\footnote{Here one could even include the metric in the field space and integration measure, as long as one supplements equation (\ref{eq:generaleff}) with a properly chosen gauge fixing term.}
\begin{equation}\label{eq:generaleff}
\Gamma[\psi,g^{\mu\nu}]=\f{i}{2}\text{Tr log}\f{\delta^2S[\psi,g^{\mu\nu}]}{\delta\psi^A\delta\psi^B}
\end{equation}
where we assume that the second derivative of the action can be (formally) parametrized as
\begin{equation}\label{eq:op}
\f{\delta^2S[\psi]}{\delta\psi^A\delta\psi^B}=C^{\mu\nu}_{A B}\nabla_\mu\nabla_\nu +2\Gamma^\mu_{A B}\nabla_\mu+W_{A B}.
\end{equation}
In order for the procedure described in section \ref{sec:eff1scalar} to be applicable, the operator in (\ref{eq:op}) must be expressible as
\begin{equation}\label{eq:op2}
\f{\delta^2S[\psi]}{\delta\psi^A\delta\psi^B}\rightarrow g^{\mu\nu}\delta_A^B\mathcal{D}_\mu\mathcal{D}_\nu +P_A^B,
\end{equation}
where $\mathcal{D}_\mu$ is the covariant derivative in some yet undetermined general space. This form can only be acchieved if the matrix $C^{\mu\nu}_{A B}$ is trivial in coordinate space i.e. $C^{\mu\nu}_{A B}=\tilde{C}_{A B}g^{\mu\nu}$. If this is the case, we arrive at (\ref{eq:op2}) by redefining the covariant derivative to absorb the linear term in $\nabla_\mu$ in (\ref{eq:op}) and multiplying the entire operator with the inverse of $C^{\mu\nu}_{A B}$. Since the matrix in front of the kinetic part commutes with the rest of the operator in (\ref{eq:op2}), the term $P^A_B$ can be interpreted as a mass matrix and the eigenvalues will be of the form $kinetic$ $term$ $+$ $mass$ $eigenvalue$. Hence, the main difficulty will be involved in finding the eigenvalues of the mass matrix. Again, the generality of the above discussion extends beyond the coupled scalar field case.
An example of its use in a more general setting can be found in \cite{Barvinsky:2009fy}

Guided by the above discussion, we can straightforwardly derive the one-loop effective action for our two-field theory. The one-loop contribution will be the sum of the eigenvalues 
\begin{equation}
\Gamma^{(1)}[{\phi},{\sigma},g^{\mu\nu}]=\f{i}{2}\text{Tr log}~\Big[\Box+\lambda_+\Big]+\f{i}{2}\text{Tr log}~\Big[\Box+\lambda_-\Big],
\end{equation}
where $\lambda_\pm$ are the eigenvalues of the mass matrix in (\ref{eq:op2}), which for our theory can directly be read off the Lagrangian (\ref{eq:matteraction})
\begin{align}
P^B_A=
\begin{pmatrix}
  m^2_\phi+(1-\xi_\phi)\f{R}{6}+\f{g{\sigma}^2}{2}+\f{\lambda_\phi{\phi}^2}{2}& \f{g}{2}{\sigma}{\phi} \\
  \f{g}{2}{\sigma}{\phi}  &m^2_\sigma+(1-\xi_\sigma)\f{R}{6}+\f{g{\phi}^2}{2}+\f{\lambda_\sigma{\sigma}^2}{2}
\end{pmatrix}.
\end{align}
We parametrize the eigenvalues as
\begin{equation}
\lambda_\pm\equiv\bigg\{\frac{m_{\phi}^2+m_{\sigma}^2}{2}+\frac{1}{6} R\bigg(1-\frac{\xi_{\phi}+\xi_{\sigma}}{2} \bigg)+\f{\Phi^2_\pm}{2}\bigg\},
\end{equation}
where 
$\Phi_\pm$ is defined in (\ref{eq:eigenf}). Retracing the steps that led to (\ref{aa2}) allows us to write the full effective action for two scalar fields to one-loop order as
\begin{align}\label{eff2}
\Gamma[{\phi},{\sigma},g^{\mu\nu}]&=S_m[{\phi},{\sigma},g^{\mu\nu}]+S_g[g^{\mu\nu}]+\delta S[{\phi},{\sigma},g^{\mu\nu}]\nonumber \\&+\int d^nx \sqrt{-g}~\f{1}{64\pi^2}\bigg\{-M_+^4\bigg[\log\bigg(\f{M^2_+}{\tilde{\mu}^2}\bigg)-\f{3}{2}\bigg]-2a_2^+(x,x)\log\bigg(\f{M_+^2}{\tilde{\mu}^2}\bigg)\bigg\}\nonumber \\&+\int d^nx \sqrt{-g}~\f{1}{64\pi^2}\bigg\{-M_-^4\bigg[\log\bigg(\f{M_-^2}{\tilde{\mu}^2}\bigg)-\f{3}{2}\bigg]-2a^-_2(x,x)\log\bigg(\f{M_-^2}{\tilde{\mu}^2}\bigg)\bigg\},
\end{align}
with the natural definitions
\begin{align}
M^2_\pm&=\frac{m_{\phi}^2+m_{\sigma}^2}{2}+\f{1}{2}\Phi^2_\pm-\f{\xi_{\phi}+\xi_{\sigma}}{12}R\\ 
a_2^\pm(x,x)&=-\f{1}{12}\Box \Phi^2_\pm+\left(5\frac{\xi_{\phi}+\xi_{\sigma}}{2}+1\right)\f{\Box R}{180}+\f{3C^2-G}{360}.
\end{align}
From the expression for the eigenvalues $\Phi_\pm$ in (\ref{eq:eigenf}) it can be determined that at the limit $g\rightarrow 0$ the result in (\ref{eff2}) reduces to two decoupled effective actions of the form (\ref{aa2}). At this stage our expression is still divergent in the limit $n\rightarrow 4$ since we have not yet renormalized, only regularized (with the scale $\tilde{\mu}$ still present). However as promised after (\ref{arbscale}), we first make some comments on the level of our approximation.

\subsection{Truncation of the effective action}
\label{sec:truncation}

In our derivation for the effective action (\ref{eff2}) we truncated under the assumption
\begin{equation}
\f{a_3(x,x)}{M^2_\pm}\ll a_2(x,x),
\end{equation}
where the terms in $a_3$ can be read e.g. from \cite{Jack:1985mw}. Schematically, for a $\varphi^4$ theory, the neglected terms are of the type
\begin{gather}
\nabla^\mu R\f{\nabla_\mu R}{M^2},\quad \nabla^\mu (\lambda\varphi^2)\f{\nabla_\mu (\lambda\varphi^2)}{M^2},\quad \nabla^\mu (\lambda\varphi^2)\f{\nabla_\mu R}{M^2},\quad R\f{\Box R}{M^2},\nonumber \\\quad R\f{\Box (\lambda\varphi^2)}{M^2},\quad (\lambda\varphi^2)\f{\Box R}{M^2},\quad (\lambda\varphi^2)\f{\Box \lambda\varphi^2}{M^2},\quad\f{\Box^2(\lambda \varphi^2)}{M^2},\quad \f{\Box^2R}{M^2},\quad\f{R^3}{M^2}.\label{apprott}
\end{gather}
It is also worth noting that terms of type
\begin{equation}
\nabla_\mu R\nabla^\mu\log\bigg(\f{M^2}{m^2}\bigg), \qquad  R\Box\log\bigg(\f{M^2}{m^2}\bigg),
\end{equation}
appear in the calculation (for instance from $a_2$ after partial integration) and are of the same order as those in (\ref{apprott}). These are also neglected.

For our two-field model (\ref{eff2}), the following terms are similarly neglected
\begin{gather}
\nabla^\mu R\f{\nabla_\mu R}{M^2_\pm},\quad \nabla^\mu \Phi^2_\pm\f{\nabla_\mu \Phi^2_\pm}{M^2_\pm},\quad \nabla^\mu \Phi^2_\pm\f{\nabla_\mu R}{M^2_\pm},\quad R\f{\Box R}{M^2_\pm},\nonumber \\ R\f{\Box \Phi^2_\pm}{M^2_\pm},\quad \Phi^2_\pm\f{\Box R}{M^2_\pm},\quad \Phi^2_\pm\f{\Box\Phi^2_\pm}{M^2_\pm},\quad\f{\Box^2\Phi^2_\pm}{M^2_\pm},\quad\f{\Box^2R}{M^2_\pm},\quad\f{R^3}{M^2_\pm},
\end{gather}
For the specific choices $\xi_\phi=\xi_\sigma=1$, $\lambda_\phi=\lambda_\sigma=0$, we have\footnote{Actually, the eigenvalues $\Phi_\pm$ in (\ref{eq:eigenf}) acquire an absolute value coming from the square root of a square, but its sign will never become explicit in the effective action (\ref{main2}).}
\begin{equation}
\Phi^2_\pm=m^2_\sigma-m^2_\phi+g\phi^2,\qquad  m^2_\phi-m^2_\sigma+g\sigma^2,
\end{equation}
and
\begin{equation}
M^2_\pm=m^2_\sigma-\f{R}{6}+\f{g\phi^2}{2},\quad\quad  m^2_\phi-\f{R}{6}+\f{g\sigma^2}{2}.
\end{equation}
From these considerations we can deduce that our approximations are valid as long our fields ($\sigma$, $\phi$ and $g^{\mu\nu}$) are slowly varying in agreement with our motivation for the anzats (\ref{HK}).

\subsection{Renormalization}
\label{sec:renormalization}

The renormalization procedure for the effective action is well-established, for example in the original paper \cite{Coleman:1973jx}, and for curved space \cite{Kirsten:1993jn,Hu:1984js}. First, we briefly present the renormalization procedure for the $\varphi^4$ theory in unbounded Minkowski space-time for a constant field. The effective action in (\ref{aa2}) becomes
\begin{gather}
\int d^4x\sqrt{-g}~\mathcal{L}_{eff}\equiv-\int d^4x\sqrt{-g}~V(\varphi)\nonumber \\=\int d^4x\sqrt{-g}~\bigg\{\bigg[-\frac{m^2}{2}\varphi^2-\f{\lambda}{4!}\varphi^4+\Lambda-\frac{\delta m^2}{2}\varphi^2-\f{\delta \lambda}{4!}\varphi^4+\delta \Lambda\nonumber \\-\f{1}{64\pi^2}\bigg(m^2+\lambda\f{\varphi^2}{2}\bigg)^2\bigg[\log\bigg(\f{m^2+\lambda\f{\varphi}{2}}{\tilde{\mu}^2}\bigg)-\f{3}{2}\bigg]\bigg\}.
\end{gather}
The counter terms $\delta m^2$, $\delta \lambda$ and $\delta \Lambda$ are fixed by a choice of renormalization conditions, for which we take
\begin{align}
m^2&=\f{\partial^2V(\varphi)}{\partial\varphi^2}\bigg\vert_{\varphi=\varphi_1},&\lambda&=\f{\partial^4V(\varphi)}{\partial\varphi^4}\bigg\vert_{\varphi=\varphi_2},&-\Lambda&=V(\varphi)\Big\vert_{\varphi=\varphi_3},
\end{align}
where the fields $\varphi_n$ are arbitrary constant renormalization points. If we choose $\varphi_1=\varphi_2=\varphi_3=0$  and further that $\Lambda =0$, we get the standard result for the effective potential \cite{Cheng:1985bj}
\begin{equation}
\label{eq:effact4}
V(\varphi)=\frac{m^2}{2}\varphi^2+\f{\lambda}{4!}\varphi^4+\f{1}{64\pi^2}\bigg[\bigg(m^2+\lambda\f{\varphi}{2}\bigg)^2\log\bigg(\f{m^2+\lambda\f{\varphi}{2}}{m^2}\bigg)-\f{\lambda\varphi^2\big(4m^2+3\lambda\varphi^2\big)}{8}\bigg].
\end{equation}
In a similar fashion, we can renormalize our two field effective action (\ref{eff2}) in a general spacetime with the following renormalization conditions\footnote{There is no need for a condition for $\delta Z$ since the one-loop contribution has no momentum dependence. This will appear at higher loop order.}
\begin{align}\label{renn1}
-m^2_\phi&=\f{\partial^2\mathcal{L}_{eff}}{\partial\phi^2},&-m^2_\sigma&=\f{\partial^2\mathcal{L}_{eff}}{\partial\sigma^2},&-\lambda_\sigma&=\f{\partial^4\mathcal{L}_{eff}}{\partial\sigma^4},\nonumber \\-\lambda_\phi&=\f{\partial^4\mathcal{L}_{eff}}{\partial\phi^4}&-g&=\f{\partial^4\mathcal{L}_{eff}}{\partial\phi^2\sigma^2}&-\f{1-\xi_\phi}{6}&=\f{\partial^3\mathcal{L}_{eff}}{\partial\phi^2\partial R},\nonumber\\-\f{1-\xi_\sigma}{6}&=\f{\partial^3\mathcal{L}_{eff}}{\partial\sigma^2\partial R},&\Lambda&=\mathcal{L}_{eff},&\alpha&=\f{\partial\mathcal{L}_{eff}}{\partial R},\nonumber \\2\beta&=\f{\partial^2\mathcal{L}_{eff}}{\partial R^2},&\epsilon_{1}&=\f{\partial\mathcal{L}_{eff}}{\partial C^2},&\epsilon_{2}&=\f{\partial\mathcal{L}_{eff}}{\partial G},\nonumber \\\eta_\phi&=\f{\partial\mathcal{L}_{eff}}{\partial\Box\phi^2},&\eta_\sigma&=\f{\partial\mathcal{L}_{eff}}{\partial\Box\sigma^2},&\kappa&=\f{\partial\mathcal{L}_{eff}}{\partial \Box R},
\end{align}
where it is understood that for every renormalized constant the renormalization point for $g^{\mu\nu}$ is $\eta^{\mu\nu}$ and for $\phi$ and $\sigma$ it is chosen to be zero. Furthermore, we will not have to deal with the infrared singularity mentioned in \cite{Coleman:1973jx} resulting from the zero scale renormalization of the four-point functions, since we are not interested in the massless limit. Also besides renormalization, the massless limit would pose additional complications in curved space due to the lack of an exponential damping factor in the heat kernel anzats (\ref{HK}). The massless limit would probably require the use of the non-local heat kernel expansion as pointed out in \cite{Barvinsky:1994ic}. One should therefore be very careful when taking the small mass limit of (\ref{eq:effact4}) (as was done in \cite{Enqvist:2011jf}), or when imposing renormalization conditions. Results for the counter terms defined in (\ref{renn1}) can be found in Appendix \ref{sec:CT} and by inserting the expressions into (\ref{eff2}) and taking the limit $n\rightarrow 4$ we get the full results for the effective action
\ba
\label{eq:fullres1}
\mathcal{L}_{eff}^{(0)}&=&
-\frac{1}{2}g^{\mu\nu}\partial_\mu\phi\partial_\nu\phi+\eta_\phi\Box\phi^2-\frac{m_\phi^2}{2}\phi^2-\f{1}{2}(1-\xi_\phi)\f{R}{6}\phi^2-\f{\lambda_\phi\phi^4}{4!}\nonumber \\
&&
-\frac{1}{2}g^{\mu\nu}\partial_\mu\sigma\partial_\nu\sigma+\eta_\sigma\Box\sigma^2-\frac{m_\sigma^2}{2}\sigma^2-\f{1}{2}(1-\xi_\sigma)\f{R}{6}\sigma^2-\f{\lambda_\sigma \sigma^4}{4!}-\f{g\phi^2\sigma^2}{4}\nonumber \\
&&+\Lambda+\alpha R+\beta  R^2+C^2 \epsilon_1+G \epsilon_2+\kappa\Box R,
\ea
and
\ba
\label{eq:fullres2}
\mathcal{L}_{eff}^{(1)}&=&\f{1}{64\pi^2}\bigg\{\f{1}{24}\Big[6 g \left(g+3 \left(\lambda _{\sigma }+\lambda _{\phi }\right)\right)\sigma ^2 \phi ^2 + \left(\xi _{\sigma }^2+\xi _{\phi }^2\right)R^2-4  \left(\xi _{\sigma } m_{\sigma }^2+m_{\phi }^2 \xi _{\phi }\right)R\nonumber\\
&&
+9 \left(g^2+\lambda _{\phi }^2\right) \phi ^4+12 \left(g m_{\sigma }^2+m_{\phi }^2 \lambda _{\phi }\right) \phi ^2-6  \left(g \xi _{\sigma }+\lambda _{\phi } \xi _{\phi }\right)R \phi^2\nonumber \\
&&
+9  \left(g^2+\lambda _{\sigma }^2\right)\sigma^4+12 \left(gm_\phi^2+ m_{\sigma }^2\lambda _{\sigma }\right)\sigma^2-6  \left(g \xi _{\phi}+\lambda _{\sigma } \xi _{\sigma }\right)R \sigma ^2
\Big]\nonumber \\
&&
- \bigg[-\frac{(G-3 C^2)}{360} -\frac{(g \square \sigma ^2+\lambda _{\phi }\square\phi ^2 )}{12} +\frac{\square R (5 \xi _{\phi }+1)}{180} +\frac{g^2 \sigma ^2 \phi ^2 m_{\phi }^2}{4 (m_{\phi }^2-m_{\sigma }^2)}
\nonumber \\
&&\phantom{\f{1}{64\pi}\f{1}{24}\Big[}\qquad\qquad\qquad\qquad
+\f{1}{2}\left(m_{\phi }^2+\frac{\phi ^2 \lambda _{\phi }}{2}+\frac{g \sigma ^2}{2}-\frac{R \xi _{\phi }}{6}\right)^2\bigg]\log \bigg(\frac{M_-^2 M_+^2}{m_{\phi }^4}\bigg)\nonumber \\
&&
- \bigg[-\frac{(G-3 C^2)}{360} -\frac{(g \square \phi ^2+\lambda _{\sigma }\square\sigma ^2 )}{12} +\frac{\square R (5 \xi _{\sigma }+1)}{180} -\frac{g^2 \sigma ^2 \phi ^2 m_{\sigma }^2}{4 (m_{\phi }^2-m_{\sigma }^2)}
\nonumber \\
&&\phantom{\f{1}{64\pi}\f{1}{24}\Big[}\qquad\qquad\qquad\qquad
+\f{1}{2}\left(m_{\sigma }^2+\frac{\sigma ^2 \lambda _{\sigma }}{2}+\frac{g \phi ^2}{2}-\frac{R \xi _{\sigma }}{6}\right)^2\bigg]\log \bigg(\frac{M_-^2 M_+^2}{m_{\sigma }^4}\bigg)
\nonumber \\
&&
+\frac{1}{12} \log \bigg(\frac{M_-^2}{M_+^2}\bigg) \Big[3  \sigma ^2(g+\lambda_\sigma)+3 \phi ^2(g+\lambda_\phi)\nonumber \\
&&
+6 m_{\sigma }^2+6 m_{\phi }^2- R(\xi_\sigma+\xi_\phi) -2 \square 
\Big]
(M_+-M_-)\bigg\}.
\ea
If we are in an unbounded space, we can use partial integration to remove the terms with a $\Box$ since they are either total divergences or beyond our approximation, cf. section \ref{sec:truncation}.

\subsection{The quantum Friedmann equation to order $R^2$}
\label{sec:Friedmann}

Here we briefly go through the steps necessary for deriving the quantum Friedmann equation used in section \ref{sec:friedmann}.
The calculation of the Einstein equation
\begin{equation}
\frac{\delta\Gamma[g^{\mu\nu},\phi,\sigma]}{\delta g^{\mu\nu}}=\f{\delta}{\delta g^{\mu\nu}}\int d^4 x\sqrt{-g} \mathcal{L}_{eff}=0,
\end{equation}
for the action (\ref{unbar}) is straightforward. Because of our truncation of the series in (\ref{series expansion}), there is no loss of accuracy if we first expand to $\mathcal{O}(R^2)$ 
and then calculate the variation w.r.t. $g^{\mu\nu}$ by using the formulae in appendix \ref{sec:variations} discarding higher-order terms according to section \ref{sec:truncation}. This gives
\begin{align}
&-\f{g_{\mu\nu}}{2}\bigg\{-\f{\partial_\rho\sigma\partial^\rho\sigma}{2}-\f{m^2_\sigma}{2}\sigma^2+\Lambda\bigg\}-\f{\partial_\mu\sigma\partial_\nu\sigma}{2}\nonumber \\&-\f{g_{\mu\nu}}{2}\bigg\{\f{1}{64\pi^2}\bigg[\f{g\sigma^2}{8}\big(4m^2_\phi+3g\sigma^2\big)-\bigg(m^2_\phi+\f{g\sigma^2}{2}\bigg)^2\log\bigg(1+\f{g\sigma^2}{2m^2_\phi}\bigg)\bigg]\bigg\}\nonumber \\&+\Big[-\f{1}{2} Rg_{\mu\nu}+R_{\mu\nu}-\nabla_\mu\nabla_\nu+g_{\mu\nu}\Box\Big]\bigg\{\alpha+\f{1}{64\pi^2}\bigg[\nonumber \\&-\f{g\sigma^2}{6}+\f{1}{3}\bigg(m^2_\phi+\f{g\sigma^2}{2}\bigg)\log\bigg(1+\f{g\sigma^2}{2m^2_\phi}\bigg)\bigg]\bigg\}\nonumber \\&-\f{3^{(1)}H_{\mu\nu}+6^{(2)}H_{\mu\nu}}{180}\f{1}{64\pi^2}\log\bigg(1+\f{g\sigma^2}{2m^2_\phi}\bigg)=0.
\end{align}
Using the tensor expressions from appendix \ref{sec:tensors} we get the Friedmann equations for the "$00$" component
\begin{align}\label{ein11}
&~~~~\f{1}{2}\bigg(\Lambda-\f{\dot{\sigma}^2}{2}-\f{m^2_\sigma}{2}\sigma^2 \bigg )+\f{1}{64\pi^2}\bigg[\f{g\sigma^2\big(4m^2_\phi+3g\sigma^2\big)}{16}-\f{1}{2}\big(m^2_\phi+\f{g\sigma^2}{2}\big)^2\log\bigg(1+\f{g\sigma^2}{2m^2_\phi}\bigg)\bigg]\nonumber \\&+\f{\dot{a}}{a}\bigg\{\f{g\sigma\dot{\sigma}}{64\pi^2}\log\bigg(1+\f{g\sigma^2}{2m^2_\phi}\bigg)\bigg\}\nonumber \\&-\f{\dot{a}^2}{a^2}\bigg\{-3\alpha+\f{1}{64\pi^2}\bigg[\f{g\sigma^2}{2}-\bigg(m^2_\phi+\f{g\sigma^2}{2}\bigg)\log\bigg(1+\f{g\sigma^2}{2m^2_\phi}\bigg)\bigg]\bigg\}\nonumber \\&+\bigg(\f{3}{2}\f{\dot{a}^4}{a^4}-\f{\dot{a}^2\ddot{a}}{a^3}+\f{1}{2}\f{\ddot{a}^2}{a^2}-\f{\dot{a}a^{(3)}}{a^2}\bigg)\f{1}{64\pi^2}\log\bigg(1+\f{g\sigma^2}{2m^2_\phi}\bigg)=0
\end{align}
and for the "$ii$" component
\begin{align}\label{ein22}
&-\f{1}{2}\bigg(\Lambda+\f{\dot{\sigma}^2}{2}-\f{m^2_\sigma}{2}\sigma^2\bigg )+\f{1}{64\pi^2}\bigg[-\f{g\sigma^2\big(4m^2_\phi+3g\sigma^2\big)}{16}\nonumber \\&-\f{1}{3}\f{g^2\sigma^2\dot{\sigma}^2}{m^2_\phi+\f{g\sigma^2}{2}}+\bigg(\f{1}{2}\big(m^2_\phi+\f{g\sigma^2}{2}\big)^2-\f{1}{3}\big(g\dot{\sigma}^2+g\sigma\ddot{\sigma}\big)\bigg)\log\bigg(1+\f{g\sigma^2}{2m^2_\phi}\bigg)\bigg]\nonumber \\&-\f{\dot{a}}{a}\bigg\{\f{2}{3}\f{g\sigma\dot{\sigma}}{64\pi^2}\log\bigg(1+\f{g\sigma^2}{2m^2_\phi}\bigg)\bigg\}\nonumber \\&-\f{1}{3}\f{\dot{a}^2}{a^2}\bigg\{3\alpha+\f{1}{64\pi^2}\bigg[-\f{g\sigma^2}{2}+\bigg(m^2_\phi+\f{g\sigma^2}{2}\bigg)\log\bigg(1+\f{g\sigma^2}{2m^2_\phi}\bigg)\bigg]\bigg\}\nonumber \\&-\f{2}{3}\f{\ddot{a}}{a}\bigg\{3\alpha+\f{1}{64\pi^2}\bigg[-\f{g\sigma^2}{2}+\bigg(m^2_\phi+\f{g\sigma^2}{2}\bigg)\log\bigg(1+\f{g\sigma^2}{2m^2_\phi}\bigg)\bigg]\bigg\}\nonumber \\&+\bigg(\f{1}{2}\f{\dot{a}^4}{a^4}-2\f{\dot{a}^2\ddot{a}}{a^3}+\f{1}{2}\f{\ddot{a}^2}{a^2}+\f{2}{3}\f{\dot{a}a^{(3)}}{a^2}+\f{1}{3}\f{a^{(4)}}{a}\bigg)\f{1}{64\pi^2}\log\bigg(1+\f{g\sigma^2}{2m^2_\phi}\bigg)=0.
\end{align}
If we neglect the $\mathcal{O}(R^2)$ contributions (four dots) and perform the redefinitions $\alpha\rightarrow 1/(16\pi G_N)$ and $\Lambda\rightarrow-1/(8\pi G_N)\Lambda'$ equation (\ref{ein11}) becomes (\ref{ein1}) and if we further solve for $\dot{a}^2/a^2$ from (\ref{ein11}) and substitute it to (\ref{ein22}) we get (\ref{ein2}). 

\subsection{Equations of motion to order $(R,g^2)$}
\label{sec:orderg2}

As a final exercise, we also show the result of further expanding the equations of motion and the Friedmann equations (\ref{bar1}), (\ref{ein1}) and (\ref{ein2}) to leading order in the coupling (still concentrating on our simplified model). The equations for $\sigma$ and $a$ can then be written in a very physically intuitive form 
\begin{align}
\ddot{\sigma}+3H\dot{\sigma}+m^2_\sigma\sigma=\f{g^2 R\sigma^3}{384\pi^2m^2_\phi},
\end{align}
for $\sigma$, and for $a$
\begin{align}
&\f{\dot{a}^2}{a^2}\bigg(1+\f{G_N g^2}{96\pi m_\phi^2}\sigma^4\bigg)+\f{\dot{a}}{a}\bigg(\f{\dot{\sigma}}{\sigma}\f{G_N g^2}{24\pi m_\phi^2}\sigma^4\bigg)=\f{8G_N\pi}{3}\rho+\f{\Lambda'}{3},\\
&\f{\ddot{a}}{a}\bigg(1+\f{G_N g^2}{96\pi m_\phi^2}\sigma^4\bigg)+\f{\dot{a}}{a}\bigg(\f{\dot{\sigma}}{\sigma}\f{G_N g^2}{48\pi m_\phi^2}\sigma^4\bigg)=\f{\Lambda'}{3}-\f{4\pi G_N}{3}\bigg[\rho+3p+\f{g^2\sigma^4}{64\pi^2m^2_\phi}\left(3\f{\dot{\sigma}^2}{\sigma^2}+\f{\ddot{\sigma}}{\sigma}\right)\bigg],
\end{align}
where $\rho$ and $p$ are the tree-level energy-density and pressure defined in section (\ref{sec:friedmann}).

The leading order correction is proportional to $\sigma^4$ in all three equations ($\sigma^3$ for the $\sigma$-equation corresponding to a $\sigma^4$ correction to the potential), but perhaps more surprising is that for the $\sigma$ equation it is also proportional to $R$, and hence vanishes in a radiation dominated background. On the other hand, we have seen in the previous sections that the quantum correction is insensitive to the gravitational operators. Hence the leading order term in $g$ does not dominate, simply because $R$ is in general small. We do not advocate the use of this approximation, but simply show it for illustration of the structure of the equations.

\section{Conclusion}
\label{sec:conclusions}
In the present paper, we have computed the renormalized one-loop 1PI effective action for two coupled and
self-interacting scalars in a general FRW background. We used a gradient expansion for the effective action and truncated  at $\mathcal{O}(R^3/M_\pm^2)$, 
where $M_{\pm}$ are eigenvalues of the mass matrix. We renormalized at the level of the action, using a set of renormalization conditions to fix also the finite parts of the counterterms \cite{Hu:1984js,Kirsten:1993jn,Elizalde:1994ds}. 

Subsequently specializing to a model of one field ($\sigma$) rolling under the influence of the
quantum fluctuations of the other ($\phi$), we derived the quantum corrected field
equations of motion. We also derived the quantum corrected Friedmann equations
(\ref{ein1}) and (\ref{ein2}), which are in general different from what is commonly referred to as the
``semi-classical'' approach. 

We solved the equations of motion for the case where $\sigma$ does not dominate the
energy density and the evolution of the scale factor is given. We found that the quantum effects can
be significant, but that gravitational operators can be largely neglected. This is important for
instance for scenarios involving subdominant curvatons \cite{subdom}. Our results improve on the discussion in \cite{Enqvist:2011jf}, 
and we have stressed how important it is to renormalize correctly to get the right physical result, using appropriate renormalization
conditions.

We then considered the more involved case, where an oscillating $\sigma$ dominates the energy density,
and solved the coupled system of field equation and Friedmann equations to find the
evolution of both $\sigma$ and $a(t)$. We again found that the effect of including
quantum corrections could be significant. This will be important for the reheating and
preheating process after inflation.

In a quadratic potential, a large initial value of $\sigma$ classically leads to slow-roll
inflation. We found that including quantum corrections shortens the period of inflation,
and may even prevent it from happening. This is just one instance of the general result
that quantum corrections spoil the flatness of the inflaton potential. Such effects are less 
pronounced for small-field inflation models.

Our general result (\ref{main1}) and (\ref{main2}) is directly applicable also to
non-minimally coupled fields, self-interacting fields, and scenarios such as two-field
inflation, resonant preheating after inflation. One may even generalize the gravity part
by not renormalizing to the Einstein-Hilbert action, but to different values of $\alpha$,
$\beta$, $\epsilon_2$ and include a cosmological constant $\Lambda$.

The work presented here should be seen as an attempt at embedding the dynamics of the
inflaton and other fields in FRW Universes in a consistent quantum field theory framework. It is particularly
suited for when the dynamics of the mean field is the main concern, but the two point function can also be computed.
For many applications (such as non-gaussianity in the CMB and warm inflation), it is necessary to go to higher order in a 1PI expansion to capture the appropriate physics \cite{higherorder}. One may also ``resum logarithms'' into a renormalization group improvement \cite{prokopec}. Alternatively, one could use the 2PI effective action \cite{sloth,tranberg,serreau}, but the numerical effort at NLO 
and the complications with renormalization currently make this option less appealing.

There are unresolved questions to do with including also $\mathcal{O}(R^2)$ (four
derivatives) in the effective Friedmann equations \cite{fourdiv}, allowing for symmetry
breaking potentials, $m_{\sigma,\phi}^2<0$, and including odd-power terms in the
potential $\sigma\phi$, $\sigma^3$, $\phi^3$, $\phi\sigma^2$, $\sigma\phi^2$. In connection with
inflation and (p)reheating, issues related to defining particle numbers relative to some
specific vacuum persist. It is likely that both the Bunch-Davies vacuum in strict
de Sitter and the adiabatic vacuum in FRW \cite{adiabatic} need to come into play, but a smooth transition from one to the other needs to be addressed. 

From a technical point of view it would be of interest to relate the renormalization prescription used here to adiabatic regularization of the semi-classical Friedmann equation. The truncation at order $R^2$ is reminiscent of the truncation at four-derivative order in the adiabatic regularization, and certainly the fact that all divergencies are cancelled originates in this correspondence. The heat kernel approach provides a straightforward way to write quantum correction directly in terms invariant operators and their counterterms only. This does not follow quite as simply at the level of the equations of motion through adiabatic regularization, although presumably we have used one of a number of prescriptions for dealing with the finite parts of the counterterms appearing in that context (see also \cite{Sobreira:2011ep}).

Also going beyond $R^2$ or even including the term $a_3$ in
the expansion of the heat kernel in a discussion of convergence. 

The huge success of inflation and the high precision anticipated from current and planned
CMB measurement mission motivates a more detailed computation of the primordial
perturbations and the dynamics of quantum fields in the early Universe. This requires us
to go beyond free-field theory and properly include interactions in the dynamics of the ``classical'' fields, and perform correct renormalization. We believe that a very useful and intuitive approach to this is in
terms of the 1PI effective action and variational equations derived from it. 


\acknowledgments We thank Mark Hindmarsh, Kari Rummukainen and Jens O. Andersen for many useful discussions. T.M. thanks the Niels Bohr International Academy for hospitality and NordForsk for support. T.M. is supported by the Finnish Academy of Science and Letters and Academy of Finland through project number 1134018. A.T. is supported by the Carlsberg Foundation and the Discovery Center.

\appendix

\section{Heat kernel method}
\label{sec:HK}

This section follows closely \cite{ParkerToms,DeWitt:2003pm}. In order to derive a solution for the heat kernel used in section \ref{sec:computation} we first use and exponential formula to write the variation of the one-loop contribution (\ref{eq:gammas}) as
\begin{equation}
\delta\Gamma^{(1)}[\varphi,g^{\mu\nu}]=\f{i}{2}\delta G^{-1} G=-\f{1}{2}\int_0^\infty {d\tau }~\delta G^{-1}e^{-i\tau G^{-1}}=\delta\bigg[-\f{i}{2}\int_0^\infty \f{d\tau }{\tau}~\text{Tr}e^{-i\tau G^{-1}}\bigg].
\end{equation}
This implies the definition for the heat kernel
\begin{equation}
\Gamma^{(1)}[\varphi,g^{\mu\nu}]=-\f{i}{2}\int d^4x \sqrt{-g}\int_0^\infty \f{d\tau }{\tau}~e^{-i\tau G^{-1}}\equiv-\f{i}{2}\int d^4x \sqrt{-g}\int_0^\infty \f{d\tau }{\tau}~K(\tau;x,x).
\end{equation}
From the exponential form of the heat kernel the following properties can be deduced
\begin{equation}\label{eq:hkcond1}
i\f{\partial}{\partial\tau}K(\tau;x,x')= G^{-1}K(\tau;x,x')
\end{equation}
and
\begin{equation}\label{eq:hkcond2}
\lim_{\tau\rightarrow 0}K(\tau;x,x')=\f{\delta(x-x')}{\sqrt{-g}},
\end{equation}
Motivated by the flat space definition for the delta function we write the ansatz for heat kernel (in $n$ dimensions)
\begin{equation}\label{eq:HK1}
K(\tau;x,x')=\f{i}{(4\pi i \tau)^{n/2}}\Delta^{1/2}(x,x')e^{i\sigma(x,x')/2\tau}\bar{\Omega}(\tau;x,x'),
\end{equation}
where $\sigma(x,x')$ is \textit{the geodetic interval} or \textit{world function} and  $\Delta^{1/2}(x,x')$ is the  \textit{Van Vleck-Morette determinant} which can be shown to satisfy
\begin{align}\label{eq:hkrel}
&\sigma_{;\mu}{\sigma_{;}}^{\mu}=2\sigma,\quad \big(\Delta{\sigma_{;}}^{\mu} \big)\!\!~_{;\mu}=n\Delta,\quad\big[\sigma\big]=0,\quad\big[\Delta^{1/2}\big]=1,\quad\big[\sigma_{;\mu}\big]=0,\nonumber \\&\big[{\Delta^{1/2}}_{;\mu}\big]=0,\quad\big[\sigma_{;\mu\nu}\big]=g_{\mu\nu},\quad\big[{\Delta^{1/2}}_{;\mu\nu}\big]=\f{1}{6}R_{\mu\nu},\quad\big[\sigma_{\mu\nu\rho}\big]=0,\nonumber \\&\big[\Box^2\Delta^{1/2}\big]=\f{\Box R}{5}+\f{R^2}{36}+\f{R_{\alpha\beta\gamma\delta}R^{\alpha\beta\gamma\delta}}{30}-\f{R_{\alpha\beta}R^{\alpha\beta}}{30},
\end{align}
where we used the notations
\begin{equation}
f(x,x')\equiv f,\quad \lim_{x\rightarrow x'} f(x,x')=\big[f\big],\quad \nabla_\mu f(x,x')\equiv f_{;\mu}.
\end{equation}
The condition (\ref{eq:hkcond2}) implies that
\begin{equation}
\bar{\Omega}(\tau;x,x')=1+\mathcal{O}(\tau),
\end{equation}
so if we write $\bar{\Omega}$ as a power series
\begin{equation}\label{eq:omegaser}
\bar{\Omega}(\tau;x,x')=\sum_{k=0}\bar{a}_k(x,x')(i\tau)^k,
\end{equation}
we kan use (\ref{eq:hkcond1}) to write a recursion formula for all the coefficients $\bar{a}_k$. In \cite{Parker:1984dj} a new form for the Feynman propagator (and hence the anzats in (\ref {eq:HK1})) was conjectured, which effectively sums all the $R$ proportional terms in (\ref{eq:omegaser}) in closed form. In addition to making the coefficients simpler, this form would also allow one to probe some of the non-local\footnote{Not expressible as linear combinations of the generally covariant geometric objects ($R,$ $R_{\mu\nu}R^{\mu\nu}$ etc.).} aspects of the effective action, which is not possible when only a finite number of terms are included. The conjecture was proven in \cite{Jack:1985mw} and again it is worth emphazising that the proof was also extended beyond real scalar theories and thus is applicable for gauge theories and fields of higher spin. 

For an operator of the form $G^{-1}=-\Box+X$ we can sum all the $R$ and $X$ contributions by writing our ansatz (\ref{eq:HK1}) as
\begin{equation}\label{eq:HK2}
K(\tau;x,x')=\f{i}{(4\pi i \tau)^{n/2}}\Delta^{1/2}(x,x')e^{i\sigma(x,x')/2\tau-i(M^2)'\tau}{\Omega}(\tau;x,x'),
\end{equation}
where
\begin{equation}
(M^2)'=X(x')-\f{R(x')}{6},
\end{equation}
where $\Omega$ has also a power series expansion of the form (\ref{eq:omegaser}). Inserting (\ref{eq:HK2}) into (\ref{eq:hkcond1}) we get the following recursion relation for the coefficients $a_k$
\begin{equation}
{\sigma_;}^\mu a_{k;\mu}+ka_k=\Delta^{-1/2}\big(\Delta^{1/2}a_{k-1}\big)\!\!~_{;\mu}^{~~\mu}+\bigg[-\f{R'}{6}+(X'-X)\bigg]a_{k-1}.
\end{equation}
With the help of the above formula and the relations in (\ref{eq:hkrel}) we can solve for the first three coefficients
\begin{align}\label{eq:hkcoeff}
&\big[a_0\big]=1,\quad\big[a_1\big]=0\nonumber \\&\big[a_2\big]=-\f{1}{6}\Box\bigg(X-\f{R}{6}\bigg)+\f{1}{180}\bigg(\Box R+R_{\alpha\beta\gamma\delta}R^{\alpha\beta\gamma\delta}-R_{\alpha\beta}R^{\alpha\beta}\bigg).
\end{align}
Using this method beyond the first few orders becomes increasingly cumbersome and other more efficient methods have been devised \cite{Gilkey:1975iq} and \cite{Avramidi:1990ug}. It is important to realize that at no point of our calculation did we choose a particular vacuum in which to define our propagator. This means that the result for (\ref{eq:HK2}) is not sensitive to boundary condition choices, which may in some instances be significant. 
\section{Results for counter terms}
\label{sec:CT}
For the renormalization conditions in (\ref{renn1}) we get the following results:
\begin{align}
\delta m^2_\phi&=\f{1}{64\pi^2}\bigg\{2m^2_\phi\lambda_\phi\bigg[1-\log\bigg(\f{m^2_\phi}{\tilde{\mu}^2}\bigg)\bigg]+2m^2_\sigma g\bigg[1-\log\bigg(\f{m^2_\sigma}{\tilde{\mu}^2}\bigg)\bigg]\bigg\}\nonumber \\\delta m^2_\sigma&=\f{1}{64\pi^2}\bigg\{2m^2_\sigma\lambda_\sigma\bigg[1-\log\bigg(\f{m^2_\sigma}{\tilde{\mu}^2}\bigg)\bigg]+2m^2_\phi g\bigg[1-\log\bigg(\f{m^2_\phi}{\tilde{\mu}^2}\bigg)\bigg]\bigg\}\nonumber \\ \delta \lambda_\phi &=\f{1}{64\pi^2}\bigg\{-6\lambda_\phi^2\log\bigg(\f{m^2_\phi}{\tilde{\mu}^2}\bigg)-6g^2\log\bigg(\f{m^2_\sigma}{\tilde{\mu}^2}\bigg)]\bigg\}\nonumber \\ \delta \lambda_\sigma &=\f{1}{64\pi^2}\bigg\{-6\lambda_\sigma^2\log\bigg(\f{m^2_\sigma}{\tilde{\mu}^2}\bigg)-6g^2\log\bigg(\f{m^2_\phi}{\tilde{\mu}^2}\bigg)\bigg\}\nonumber \\ \delta g &=\f{1}{64\pi^2}\bigg\{\f{2 g}{m^2_\phi-m^2_\sigma}\bigg[g(m^2_\phi-m^2_\sigma)+\bigg(m^2_\sigma\lambda_\phi-m^2_\phi(g+\lambda_\phi)\bigg)\log\bigg(\f{m^2_\phi}{\tilde{\mu}^2}\bigg)\nonumber \\\phantom{\delta g} &\phantom{=\f{1}{64\pi^2}\bigg\{\f{2 g}{m^2_\phi-m^2_\sigma}\bigg[g(m^2_\phi-m^2_\sigma)}~-\bigg(m^2_\phi\lambda_\sigma-m^2_\sigma(g+\lambda_\sigma)\bigg)\log\bigg(\f{m^2_\sigma}{\tilde{\mu}^2}\bigg)\bigg]\bigg\}\nonumber \\ \delta \xi_\phi &=\f{1}{64\pi^2}\bigg\{-2\lambda_\phi\xi_\phi\log\bigg(\f{m^2_\phi}{\tilde{\mu}^2}\bigg)-2g\xi_\sigma\log \bigg(\f{m^2_\sigma}{\tilde{\mu}^2}\bigg)\bigg\}\nonumber \\ \delta \xi_\sigma &=\f{1}{64\pi^2}\bigg\{-2\lambda_\sigma\xi_\sigma\log\bigg(\f{m^2_\sigma}{\tilde{\mu}^2}\bigg) -2g\xi_\phi\log\bigg(\f{m^2_\phi}{\tilde{\mu}^2}\bigg)\bigg\}\nonumber \\ \delta\Lambda &=\f{1}{64\pi^2}\bigg\{\f{m^4_\phi}{2}\bigg[-3+2\log\bigg(\f{m^2_\phi}{\tilde{\mu}^2}\bigg)\bigg]+ \f{m^4_\sigma}{2}\bigg[-3+2\log\bigg(\f{m^2_\sigma}{\tilde{\mu}^2}\bigg)\bigg]\bigg\}\nonumber \\ \delta\alpha &=\f{1}{64\pi^2}\bigg\{\f{m^2_\phi\xi_\phi}{3}\bigg[1-\log\bigg(\f{m^2_\phi}{\tilde{\mu}^2}\bigg)\bigg]+\f{m^2_\sigma\xi_\sigma}{3}\bigg[1-\log\bigg(\f{m^2_\sigma}{\tilde{\mu}^2}\bigg)\bigg]\bigg\}\nonumber \\ \delta\beta &=\f{1}{64\pi^2}\bigg\{\f{\xi_\phi^2}{36}\log\bigg(\f{m^2_\phi}{\tilde{\mu}^2}\bigg)+\f{\xi_\sigma^2}{36}\log\bigg(\f{m^2_\sigma}{\tilde{\mu}^2}\bigg)\bigg\}\nonumber \\ \delta\epsilon_1 &=\f{1}{64\pi^2}\bigg\{\f{1}{60}\log\bigg(\f{m^2_\phi}{\tilde{\mu}^2}\bigg)+\f{1}{60}\log\bigg(\f{m^2_\sigma}{\tilde{\mu}^2}\bigg)\bigg\}\nonumber \\ \delta\epsilon_2 &=\f{1}{64\pi^2}\bigg\{-\f{1}{180}\log\bigg(\f{m^2_\phi}{\tilde{\mu}^2}\bigg)-\f{1}{180}\log\bigg(\f{m^2_\sigma}{\tilde{\mu}^2}\bigg)\bigg\}\nonumber \\ \delta\eta_\phi &=\f{1}{64\pi^2}\bigg\{-\f{\lambda_\phi}{6}\log\bigg(\f{m^2_\phi}{\tilde{\mu}^2}\bigg)-\f{g}{6}\log\bigg(\f{m^2_\sigma}{\tilde{\mu}^2}\bigg)\bigg\}\nonumber \\ \delta\eta_\sigma &=\f{1}{64\pi^2}\bigg\{-\f{\lambda_\sigma}{6}\log\bigg(\f{m^2_\sigma}{\tilde{\mu}^2}\bigg)-\f{g}{6}\log\bigg(\f{m^2_\phi}{\tilde{\mu}^2}\bigg)\bigg\}\nonumber \\ \delta\kappa &=\f{1}{64\pi^2}\bigg\{\f{1+5\xi_\phi}{90}\log\bigg(\f{m^2_\phi}{\tilde{\mu}^2}\bigg)+\f{1+5\xi_\sigma}{90}\log\bigg(\f{m^2_\sigma}{\tilde{\mu}^2}\bigg)\bigg\},
\end{align}
where $\log(\tilde{\mu}^2)$ is defined in (\ref{arbscale}). 

\section{Variational formulae}
\label{sec:variations}
In this appendix we state the needed formulae for calculating variations with respect to the metric. They can be derived with the following identities
\begin{align}
\delta g^{\mu\sigma}&=-g^{\nu\sigma}g^{\mu\rho}\delta g_{\rho\nu},\\
\delta\sqrt{-g}&=-\frac{1}{2}\sqrt{-g}g_{\mu\nu}~\delta g^{\mu\nu},\\
 \delta {R^\rho}_{\sigma\mu\nu}&=\nabla_\mu\delta\Gamma^\rho_{\nu\sigma}-\nabla_\nu\delta\Gamma^\rho_{\mu\sigma},\\ 
 \delta\Gamma^\gamma_{\alpha\beta}&=\f{g^{\gamma\rho}}{2}\big(\nabla_\alpha\delta g_{\rho\beta}+\nabla_\beta\delta g_{\rho\alpha}-\nabla_\rho\delta g_{\alpha\beta}\big),\\ 
 \sqrt{-g}\nabla_\mu A^\mu&=\partial_\mu\big(\sqrt{-g}A^\mu\big).
\end{align}
We also make use of the fact that the fields vanish at infinity and that the covariant derivative satisfies the Leibniz rule. The variations of the geometric tensors up to mass dimension four are
\begin{equation}
\f{1}{\sqrt{-g}}\f{\delta}{\delta g^{\mu\nu}}\int d^4x\sqrt{-g}~Rf(x)=\big[-\f{1}{2} Rg_{\mu\nu}+R_{\mu\nu}-\nabla_\mu\nabla_\nu+g_{\mu\nu}\Box\big]f(x)\label{var},
\end{equation}
\begin{equation}
\label{var1}\f{1}{\sqrt{-g}}\f{\delta}{\delta g^{\mu\nu}}\int d^4x\sqrt{-g}~R^2f(x)=\big[-\f{1}{2} Rg_{\mu\nu}+2R_{\mu\nu}-2\nabla_\mu\nabla_\nu+2g_{\mu\nu}\Box\big]Rf(x),
\end{equation}
\begin{gather}
\f{1}{\sqrt{-g}}\f{\delta}{\delta g^{\mu\nu}}\int d^4x\sqrt{-g}~R_{\mu\nu}R^{\mu\nu}f(x)=-\f{f(x)}{2} R_{\alpha\beta}R^{\alpha\beta}g_{\mu\nu}+2R_{\mu\rho}R^{\rho}_{~\nu}f(x)\nonumber\\+g_{\mu\nu}\nabla_\rho\nabla_\delta \big(R^{\rho\delta}f(x)\big)-2\nabla^\rho\nabla_\nu\big (R_{\rho\mu}f(x)\big)+\Box\big(R_{\mu\nu}f(x)\big),\label{var2}
\end{gather}
\begin{gather}
\f{1}{\sqrt{-g}}\f{\delta}{\delta g^{\mu\nu}}\int d^4x\sqrt{-g}~ R^{\alpha\sigma\gamma\delta}R_{\alpha\sigma\gamma\delta}f(x)=-\f{f(x)}{2}R^{\alpha\sigma\gamma\delta}R_{\alpha\sigma\gamma\delta}g_{\mu\nu}+2{R_\mu}^{\rho\alpha\sigma} R_{\nu\rho\alpha\sigma}f(x)\nonumber \\+4\nabla^\sigma\nabla^\rho\big(R_{\mu\sigma\nu\rho}f(x)\big).\label{var3}
\end{gather}
If in (\ref{var}), (\ref{var1}), (\ref{var2}) and (\ref{var3}) we choose $f(x)=1$, we can use the Bianchi identites and commutator formulae for the covariant derivative to get the standard results \cite{birreldavies}
\begin{align}
G_{\mu\nu}\equiv\f{1}{\sqrt{-g}}\f{\delta}{\delta g^{\mu\nu}}\int d^4x\sqrt{-g}~R=-\f{1}{2} Rg_{\mu\nu}+R_{\mu\nu},
\end{align}
\begin{align}
~^{(1)}H_{\mu\nu}\equiv\f{1}{\sqrt{-g}}\f{\delta}{\delta g^{\mu\nu}}\int d^4x\sqrt{-g}~R^2=-\f{1}{2} R^2g_{\mu\nu}+2R_{\mu\nu}R-2\nabla_\mu\nabla_\nu R+2g_{\mu\nu}\Box R,
\end{align}
\begin{align}
~^{(2)}H_{\mu\nu}&\equiv\f{1}{\sqrt{-g}}\f{\delta}{\delta g^{\mu\nu}}\int d^4x\sqrt{-g}~R^{\mu\nu}R_{\mu\nu}\nonumber \\&=-\f{1}{2}R_{\alpha\beta}R^{\alpha\beta}g_{\mu\nu}+ 2R_{\rho\nu\gamma\mu}R^{\rho\gamma}-\nabla_\nu\nabla_\mu R+\f{1}{2}\Box R g_{\mu\nu}+\Box R_{\mu\nu},
\end{align}
and
\begin{align}
H_{\mu\nu}&\equiv\f{1}{\sqrt{-g}}\f{\delta}{\delta g^{\mu\nu}}\int d^4x\sqrt{-g}~R^{\mu\nu\sigma\delta}R_{\mu\nu\sigma\delta}\nonumber \\&=
-\f{g_{\mu\nu}}{2}R^{\alpha\sigma\gamma\delta}R_{\alpha\sigma\gamma\delta} +2{R_\mu}^{\rho\alpha\sigma}R_{\nu\rho\alpha\sigma}+4R_{\sigma\mu\gamma\nu}R^{\gamma\sigma}- 4R_{\mu\gamma}{R^\gamma}_{\nu}+4\Box R_{\mu\nu}-2\nabla_\mu\nabla_\nu R.
\end{align}
In four dimensions the higher order tensors are connected via the Gauss-Bonnet theorem
\begin{gather}
\f{1}{\sqrt{-g}}\f{\delta}{\delta g^{\mu\nu}}\int d^4x\sqrt{-g}~ \big(R^{\mu\nu\sigma\delta}R_{\mu\nu\sigma\delta}+R^2-4R^{\mu\nu}R_{\mu\nu}\big)=0,\\ \Leftrightarrow H_{\mu\nu}=-^{(1)}H_{\mu\nu}+4^{(2)}H_{\mu\nu}\label{GB}.
\end{gather}

\section{Geometric tensors in FRW}
\label{sec:tensors}

When we choose our metric to be 
\begin{equation}
g_{\mu\nu}dx^\mu dx^\nu=-dt^2+a(t)^2d\mathbf{x}^2,
\end{equation}
i.e. of the Friedmann-Robertson-Walker type, we will make frequent use of the following tensors
\begin{align}\label{ten}
R&=6\bigg(\f{\dot{a}^2}{a^2}+\f{\ddot{a}}{a}\bigg),\quad C^2=0,\quad G=24\f{\dot{a}^2\dda}{a^3},\\ 
R_{00}&=-3\f{\ddot{a}}{a},\quad R_{ii}=a^2\bigg(2\f{\dot{a}^2}{a^2}+\f{\ddot{a}}{a}\bigg),\\ 
\big(-\nabla_0\nabla_0+g_{00}\Box\big)f(t)&=3\f{\dot{a}}{a}\partial_0f(t),\\ \big(-\nabla_i\nabla_i+g_{ii}\Box\big)f(t)&=-a^2\bigg(2\f{\dot{a}}{a}\partial_0+\partial_0^2\bigg)f(t),\\ ~^{(1)}H_{00}&=-54\f{\dot{a}^4}{a^4}+36\f{\dot{a}^2\ddot{a}}{a^3}-18\f{\ddot{a}^2}{a^2}+36\f{\dot{a}a^{(3)}}{a^2},\\
 ~^{(1)}H_{ii}&=a^2\bigg(-18\f{\dot{a}^4}{a^4}+72\f{\dot{a}^2\ddot{a}}{a^3}-18\f{\ddot{a}^2}{a^2}-24\f{\dot{a}a^{(3)}}{a^2}-12\f{a^{(4)}}{a}\bigg),\\ 
 ~^{(2)}H_{00}&=-18\f{\dot{a}^4}{a^4}+12\f{\dot{a}^2\ddot{a}}{a^3}-6\f{\ddot{a}^2}{a^2}+12\f{\dot{a}a^{(3)}}{a^2},\\ 
 ~^{(2)}H_{ii}&=a^2\bigg(-6\f{\dot{a}^4}{a^4}+24\f{\dot{a}^2\ddot{a}}{a^3}-6\f{\ddot{a}^2}{a^2}-8\f{\dot{a}a^{(3)}}{a^2}-4\f{a^{(4)}}{a}\bigg).
\end{align}
Having the Gauss-Bonnet theorem (\ref{GB}) means that we will not have any need for an expression for $H_{\mu\nu}$.


\end{document}